\pdfoutput=1 
\documentclass[
aip, 
jcp, 
reprint,
floatfix,
amsmath,amssymb,]{revtex4-1}

\usepackage{graphicx}
\usepackage[utf8]{inputenc}
\usepackage[T1]{fontenc}
\usepackage{mathptmx}
\usepackage{etoolbox}
\usepackage{hyperref}
\usepackage{bm}

\draft 

\makeatletter
\def\@email#1#2{%
 \endgroup
 \patchcmd{\titleblock@produce}
  {\frontmatter@RRAPformat}
  {\frontmatter@RRAPformat{\produce@RRAP{*#1\href{mailto:#2}{#2}}}\frontmatter@RRAPformat}
  {}{}
}%
\makeatother

\begin{document}

\title{A Hessian-Based Assessment of Atomic Forces for Training Machine Learning Interatomic Potentials}

\author{Marius Herbold}
\author{J\"{o}rg Behler}
\email{joerg.behler@uni-goettingen.de}
\affiliation{Universit\"{a}t G\"{o}ttingen, Institut f\"{u}r Physikalische Chemie, Theoretische Chemie, Tammannstra\ss{}e 6, 37077 G\"{o}ttingen, Germany}

\date{\today}

\begin{abstract}
In recent years, many types of machine learning potentials (MLPs) have been introduced, which are able to represent high-dimensional potential-energy surfaces (PES) with close to first-principles accuracy. Most  current MLPs rely on atomic energy contributions given as a function of the local chemical environments. Frequently, in addition to total energies also atomic forces are used to construct the potentials, as they provide detailed local information about the PES. Since many systems are too large for electronic structure calculations, obtaining reliable reference forces from smaller subsystems like molecular fragments or clusters can substantially simplify the construction of the training sets. Here, we propose a method to determine structurally converged molecular fragments providing reliable atomic forces based on an analysis of the Hessian. The method, which serves as a locality test and allows to estimate the importance of long-range interactions, is illustrated for a series of molecular model systems and the metal-organic framework MOF-5 as an example for a complex organic-inorganic hybrid material.
\end{abstract}


\maketitle 

\section{Introduction}\label{sec:introduction}

Despite the ever increasing computational power, large-scale atomistic simulations of complex systems remain a challenging task. In many fields of research like chemistry, molecular biology, materials science and physics interatomic potentials are playing a crucial role, as they determine the accuracy and the quality of the simulation results. While electronic structure methods like the density functional theory (DFT) provide a reliable and accurate description of many systems, the high computational costs of DFT, when directly used in simulations like ab initio molecular dynamics (MD) or Monte Carlo (MC), severely limit the accessible system size and the number of simulation steps that can be performed. Thus, more efficient atomistic potentials are needed providing a direct functional relation between the atomic structure and the potential energy. The majority of such potential energy surfaces (PES) available in the literature relies on physics-based approximations, and the employed functional forms pose an intrinsic limitation for the accuracy that can be reached.

In recent years, machine learning potentials (MLPs)~\cite{MH0873,MH0875,P5788,MH0877}, which combine the efficiency of simple empirical potentials and the accuracy of electronic structure methods, have become a promising alternative approach to represent the PES, with a variety of successful examples like neural network potentials (NNPs) \cite{JB10,MH011,MH08710,MH08711,MH08712,P5817}, Gaussian approximation potentials (GAPs) \cite{P2630,MH08241}, moment tensor potentials (MTPs) \cite{MH08714}, spectral neighbor analysis potentials (SNAPs)\cite{MH08715}, atomic cluster expansion (ACE)~\cite{MH08716} and many others \cite{MH08717, MH08242}. MLPs rely on very flexible functional forms and nowadays allow to represent even complicated high-dimensional PESs of systems containing thousands of atoms with excellent accuracy. 

Starting with the introduction of high-dimensional neural network potentials (HDNNPs)~\cite{P1174} as a first MLP of the second generation~\cite{P6018,P5977} in 2007, machine learning potentials have become applicable to large systems by expressing the total energy of the system, or at least a part of it, as a sum of environment-dependent atomic energies $E_i$,
\begin{eqnarray}
E=\sum_{i=1}^{N_{\mathrm{atom}}} E_i \quad .\label{eq:etot}
\end{eqnarray}
Here, the atomic energies are computed by machine learning algorithms using a set of suitable descriptors serving as local structural fingerprints that may be predefined \cite{P2882,P3885,P5398,P5658} or learned~\cite{P4937,MH08711,MH08710}.
In addition, also long-range electrostatic interactions can be included, e.g. in third-generation MLPs, which are based on local environment-dependent atomic partial charges determined by machine learning \cite{P2041,P2962,P5629,MH08711,P5313,P5614}, or in fourth-generation MLPs also taking non-local charge transfer into account~\cite{P4419,P5859,P5932}.

While in Eq.~\ref{eq:etot} the complexity of the full-dimensional PES is split into lower-dimensional atomic energy contributions, the atomic energies are not physical observables. Consequently, total energies need to be used for training, and the energy partitioning is implicitly done by the machine learning algorithm. 
In particular for large systems, the separation into atomic energies, which are constructed to reproduce the total energy, is not unique and subject to error compensation among the very flexible atomic energies, which are just mathematical auxiliary quantities. This reduces the transferability of the potentials. Moreover, the evaluation of the reference data set by electronic structure calculations is often the cost-determining part of MLP development, making it desirable to extract as much information as possible from the calculations. For both reasons, apart from energies also the use of atomic forces is nowadays a standard procedure in the training of MLPs~\cite{P0834,P3114,P2287,P5684}. Forces offer the advantage of being physically meaningful observables providing valuable local atomic information about the PES. Moreoever, exploiting in addition to the total energy also the 3$N_{\mathrm{atom}}$ force components can potentially reduce the required number of reference calculations.

It is important to note that energies and their negative gradients, i.e., the forces, provided by MLPs depend on the same parameters and thus are not trained independently. This does not only ensure energy conservation in applications like MD, but also has the consequence that the training process requires highly consistent energies and forces in the reference set. This makes a stringent level of convergence of the reference calculations mandatory, both in terms of the general settings of the employed electronic structure codes and in terms of system size. 

As it can be shown that Eq.~\ref{eq:etot} results in an 
effective environment-dependence of the forces corresponding up to twice the environment radius defining the atomic energies in MLPs~\cite{P5128}, sufficiently large systems have to be used.
This raises the question regarding the minimum system size required to reach a predefined degree of convergence of the force vector acting on a particular atom in the system. Due to the very different interactions in different types of molecules and materials, the answer will be highly system-dependent, calling for a general method that can be used to probe the locality of the atomic forces. With such a method at hand, the minimum system size could be determined that is required to obtained well-converged forces suitable for training and validating accurate MLPs. 

Using comparably small systems for training MLPs, which can then be applied to simulations of much larger structures, is a common procedure starting with the emergence of HDNNPs, and many successful applications from bulk materials~\cite{P3114,P5537} to molecules~\cite{P4585} have been reported. However, we note that some systems cannot be described by local energies only. In this case e.g. long-range electrostatic interactions based on local charges can be included.~\cite{P5977} For systems exhibiting non-local dependencies in the electronic structure the emerging class of fourth-generation MLPs~\cite{P4419,P5859,P5932} can be employed. In all these cases a locality test to assess the applicability of Eq.~\ref{eq:etot} is needed. 

A locality test for MLPs has been proposed by Deringer and Cs\'{a}nyi for the example of amorphous carbon~\cite{MH088}. In this approach the fluctuations of a given force vector are monitored with the local environment frozen up to a certain radius while varying atomic positions outside. The environment can then be increased until the force fluctuations remain below a predefined threshold. This method is very general, but it relies on a large number of electronic structure calculations, which must cover a representative set of different atomic configurations outside the local atomic environments.

In this work, we propose an alternative locality test, which is analytic and based on the Hessian, i.e. the second derivative of the potential energy with respect to the atomic positions. This second energy derivative is equivalent to the first derivative of the energy gradient and thus provides direct information about the dependence of the forces on the atomic positions in the system. The analysis of the Hessian allows us to determine the required spatial extension of the atomic environments, which can be used to define a minimum fragment radius $r_\mathrm{frag}$ needed to obtain converged forces in molecular fragments that agree - within a predefined tolerance - with forces in a much larger molecular or periodic bulk system. 

Using the Hessian offers several advantages. First, in contrast to direct convergence tests of the force vector as a function of the environment radius, the dependence of the investigated force on each individual neighboring atom can be quantified individually. This dependence can be very different and anisotropic depending on the structure and chemical interactions. Further, the calculation of the Hessian is a well-defined procedure that ensures a systematic investigation of the role of all atoms in the system without the need to rely on converged statistical sampling of neighboring atomic positions.  
Moreover, forces can be affected by the cancellation of contributions arising from the symmetry of the system, which is not the case for Hessian matrix elements. For instance, in crystalline environments the forces are small irrespective of the environment radius  pretending an artificial early convergence with system size. For these reasons, monitoring the convergence of forces as a function of environment radius can be very challenging.

The Hessian-based locality test we propose in this work is first illustrated for a series of simple one-dimensional model systems with different types of bonding affecting the range of interactions. We then apply the method to the metal-organic framework (MOF) MOF-5, also known as IRMOF-1 \cite{MH001,MH078}, which is a challenging benchmark not only because of its size but also because of the very complex interactions in this system. MOFs are nanoporous crystalline materials consisting of organic linker molecules and inorganic secondary building units (SBU) with a huge range of choices for the SBU and linker \cite{MH002,MH003,MH111,MH112,MH113}. Even MOFs with a combination of different SBUs or linkers have been reported \cite{MH111,MH115}, as well as postsynthetic modifications \cite{MH003,MH004,MH110},  different functionalizations \cite{MH003,MH106} and MOF composites \cite{MH107}. Because of the manifold ways to design and fine-tune the properties of MOFs, these structures are relevant for many applications like gas storage and separation, catalysis and optical devices \cite{MH003,MH007,MH105,MH106,MH109}. Theoretical investigations are of high interest in order to develop new MOFs and to analyze and predict their properties \cite{MH013}. For such theoretical studies reliable and accurate interatomic potentials are needed \cite{P6059,P6060}.

\section{Method}\label{sec:methods}

\subsection{Hessian}
For assessing the influence of all atoms in the system on all force vectors $\mathbf{f}$ we use the Hessian matrix $\mathbf{H}$ of dimension $3N_{\mathrm{atom}} \times 3N_{\mathrm{atom}}$ with elements

\begin{eqnarray}
H_{A_{\alpha} B_{\beta}}
=\frac{\partial^2 E}{\partial {A_{\alpha}} \partial {B_{\beta}}}
=-\frac{\partial f_{B_{\beta}}}{\partial {A_{\alpha}}}=-\frac{\partial f_{A_{\alpha}}}{\partial {B_{\beta}}} \quad . \label{eq:hessian_atomic_sub_matrix_norm}
\end{eqnarray}

\begin{figure}
    \includegraphics[width=\linewidth]{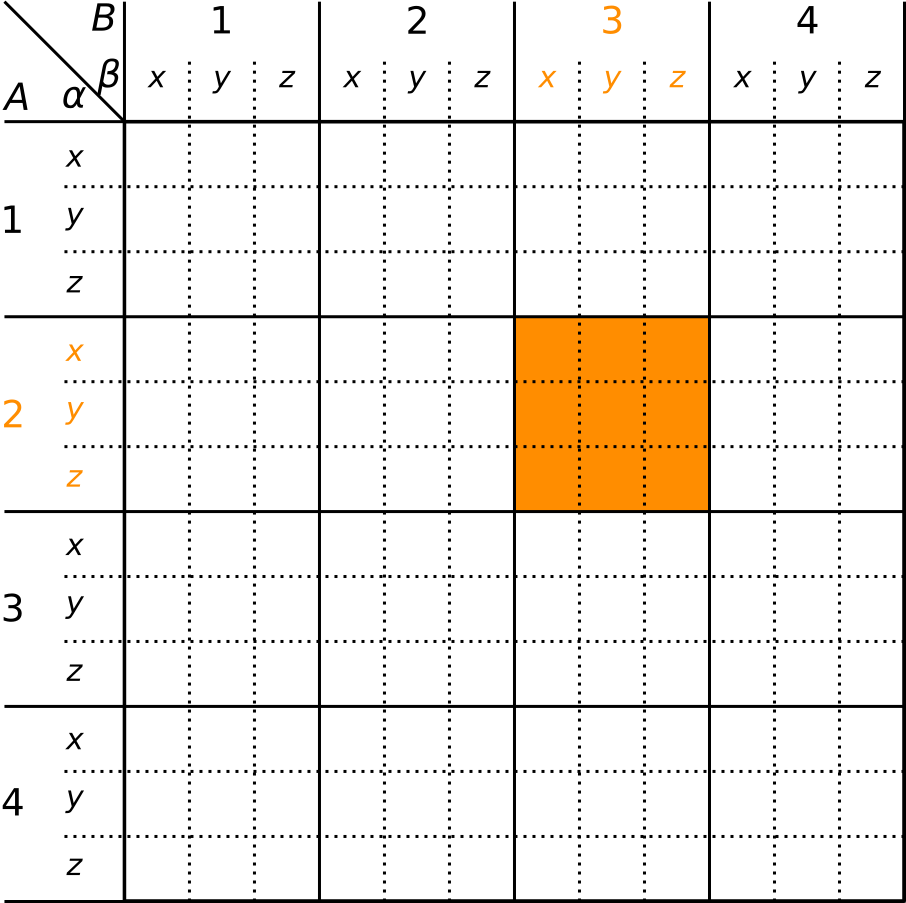}
    \caption{Structure of the Hessian matrix $\mathbf{H}$ for a system containing $N_{\mathrm{atom}} = 4$ atoms. The atomic Hessian submatrix $\mathbf{h}_\mathrm{23}$ highlighted in orange describes the interaction between atoms $A = 2$ and $B = 3$ as defined by Eq. \ref{eq:hessian_atomic_sub_matrix_norm}.}
    \label{fig:Hessian-Submatrix}
\end{figure}

$A_{\alpha}$ and $B_{\beta}$ represent Cartesian coordinates of atoms $A$ and $B$, respectively, with $\alpha,\beta=\{x,y,z\}$. The interaction between two atoms $A$ and $B$ is described by a $3 \times 3$ atomic submatrix $\mathbf{h}_{AB}$, as shown in Fig.~\ref{fig:Hessian-Submatrix}.  To represent this interaction between two atoms by a scalar quantity, we use the norm of the atomic Hessian submatrix $||\mathbf{h}_{AB}||$,

\begin{equation}
    ||\mathbf{h}_{AB}|| = \sqrt{\sum_{\alpha = {x,y,z}} \sum_{\beta = {x,y,z}} h^2_{A_{\alpha} B_{\beta}}}
     \quad .
\end{equation}

Depending on the implementation in the electronic structure code, the matrix elements $H_{A_{\alpha} B_{\beta}}$, or equivalently $h_{A_{\alpha} B_{\beta}}$, can be calculated either analytically or using finite differences.

\subsection{Hessian Group Matrix}

When constructing molecular fragments from a larger system for training machine learning potentials, it is important to investigate the convergence of the force vector $\mathbf{f}^{\mathrm{frag}}_{A}$  acting on the central atom $A$ in the fragment with respect to the force $\mathbf{f}_{A}$ in the full system as a function of the fragment radius $r_\mathrm{frag}$. In case of a periodic bulk system with an infinite number of atoms the reference force vector $\mathbf{f}_{A}$ can either be determined from the bulk system itself or from a very large fragment providing essentially the same force values. Using a very large fragment allows to analyze the role of each individual neighboring atom as a function of distance, which would not be possible when using a small periodic unit cell due to the existence of multiple periodic images of the same atom.

Starting from this large reference fragment, the question to be answered is then how much the fragment radius can be reduced for the MLP training set construction without introducing significant errors in the forces.
Decreasing the fragment radius, i.e., using a spherical environment, corresponds to simultaneously removing atoms in all spatial directions. It is therefore also of interest to quantify the combined influence of all atoms absent in the smaller fragment.  
This influence is given by the Hessian group matrix, which we define as
\begin{equation}
    \mathbf{G}^g_{A} = \sum_{B\in g} \mathbf{h}_{AB} \quad,
    \label{eq:hessian_group_sub_matrix}
\end{equation}
with the group of atoms removed from the reference fragment labeled by $g$. 
Hence, $\mathbf{G}^g_{A}$ is the sum of all atomic Hessian submatrices describing the interactions between the central atom $A$ and all atoms $B$ of the reference system beyond $r_\mathrm{frag}$ that are included in the group. 

Like in case of the individual atomic Hessian submatrices, the norm of $\mathbf{G}^g_{A}$ can be computed to describe the joint interaction of $A$ with all atoms in the group by a single number.
We note that the summation in Eq.~\ref{eq:hessian_group_sub_matrix} in principle allows contributions of different atoms $B$ to cancel each other, resulting in a system-dependent property in line with the expectation to find varying minimum fragment radii for different systems with different bonding situations. 

\begin{figure}
    \centering
    \includegraphics[width=\linewidth]{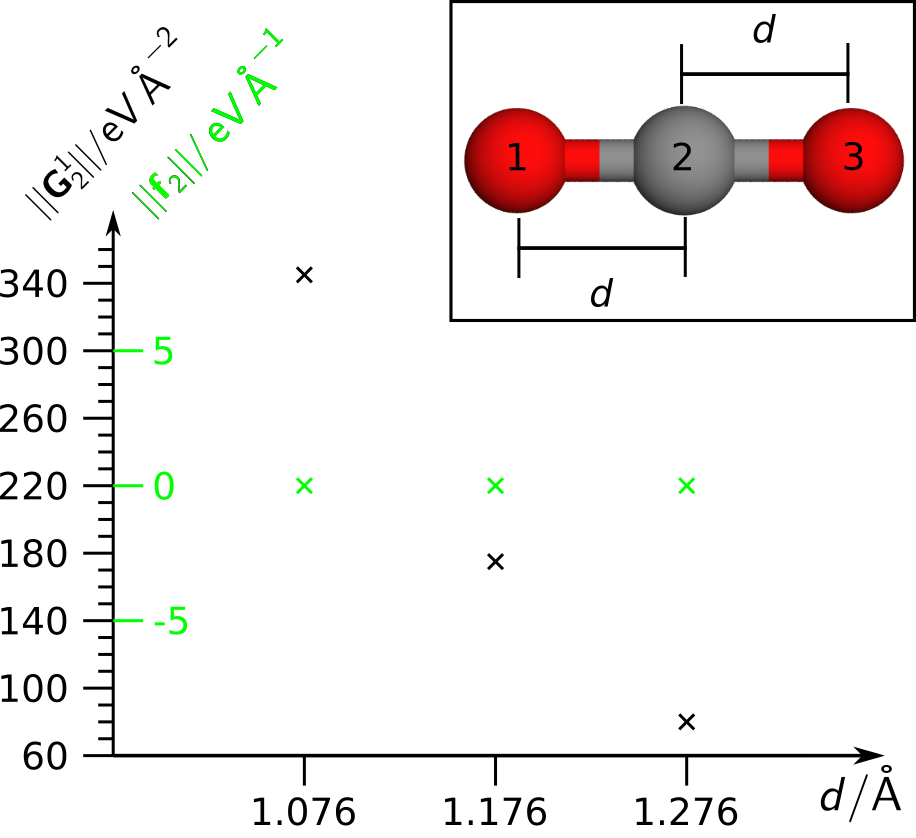}
    \caption{Force norm of the central carbon $||\mathbf{f}_2||$ (green) and Hessian group matrix norm $||\mathbf{G}_2^1||$ (black) for a CO$_2$ molecule. The CO bonds are symmetrically stretched or compressed by $d=\pm$0.1~\AA{} with respect to the DFT equilibrium bond length of $1.176\,$\AA{}. For all three geometries the force is zero because of the molecular symmetry, while the Hessian group matrix norm reflects the varying strength of the atomic interactions.}
    \label{fig:Model_System_CO2}
\end{figure}

We illustrate the concept of a Hessian group matrix using the example of a carbon dioxide molecule shown in Fig.~\ref{fig:Model_System_CO2}.
Each atom-atom interaction is described by an atomic Hessian submatrix $\mathbf{h}_{AB}$ and thus the interaction between the central carbon atom $A=2$ and the oxygen atoms $B=1,3$ in group $g=1$ is given by 

\begin{equation}
   \mathbf{G}^{\mathrm{1}}_{2} = \sum_{B\in g} \mathbf{h}_{2B} = \mathbf{h}_{21} + \mathbf{h}_{23} \quad .
\end{equation}

For a symmetric molecule, the force acting on the carbon atom is zero irrespective of the length of the CO bonds. Thus, in such symmetric situations, which are omnipresent also in systems like periodic crystals, the force has to be used with great care for the determination of the interaction range. However, as can be seen in Fig.~\ref{fig:Model_System_CO2}, the Hessian group matrix norm is nonzero for all interatomic distances even for a symmetric structure and decays with atomic separation, making this quantity a useful measure to describe the atomic interactions also in highly symmetric environments.

The CO$_2$ case is trivial in that the group contains all other atoms in the molecule apart from the central carbon atom. Generally, in very large molecular systems or bulk materials, many bonds will be broken when cutting a fragment from the system. The resulting changes in the electronic structure can strongly alter the interactions between the remaining atoms and thus need to be reduced as much as possible, which has some similarity to constructing the quantum mechanical region in QM/MM simulations. Still, some common procedures successfully employed in QM/MM \cite{MH118,P1221}, like the use of pseudopotentials or embedding the system in point charges, cannot be applied here. The reason is that by construction the interactions in MLPs are strictly local and introducing any information or assumption about the structure outside the environment radius violates this locality ansatz. As a consequence, such information would result in contradictory training data in that a force that is assumed to be local is influenced by information from outside the environment radius.

The strategy we follow here is to saturate the dangling bonds by hydrogen atoms. If, however, this saturation would be insufficient to avoid a substantial change in the electronic structure, e.g. in the case of cutting aromatic ring systems, critical functional groups will be included completely in the fragment even if parts of the functional group are outside the fragment radius $r_\mathrm{frag}$. These extended fragments are then finally also terminated by hydrogen atoms. More details about the procedure will be given in the discussion of the specific examples below. 

Still, bond breaking and subsequent saturation by hydrogen atoms is a chemical modification, which in principle can influence the force acting on the central atom in the fragment. Thus the atomic Hessian submatrices of the added hydrogen atoms $b$, i.e., $\mathbf{h}_{Ab}$, are removed from the Hessian group matrix to yield the effective Hessian group matrix

\begin{equation}
    \mathbf{G}'^{g}_{A} = \sum_{B\in \mathrm{g}} \mathbf{h}_{AB} - \sum_{b} \mathbf{h}_{Ab} \quad ,
    \label{eq:hessian_extgroup_sub_matrix}
\end{equation}

labeled by a prime, which we use in this work.
Alternatively, it would be possible not to explicitly consider the atomic Hessian submatrices of the saturating hydrogen atoms and thus to employ $\mathbf{G}^{g}_{A}$ directly.
Regardless of this choice the interactions of the central atoms with the saturating hydrogen atoms at the periphery of the molecular fragments are usually small and thus $\mathbf{G}^{g}_{A}$ and $\mathbf{G}'^{g}_{A}$ are very similar for reasonably sized fragments. 

\section{Computational Details}\label{sec:comp_details}

All DFT calculations reported in this work have been carried  out  using  the  FHI-aims  code~\cite{P2189} (release version 171221). FHI-aims is an all-electron code employing a numerical atomic orbital basis, which is determined using free atom calculations. The RPBE functional has been employed\cite{P0120} in combination with corrections to include dispersion interactions according to the method of Tkatchenko and Scheffler\cite{MH095}. ``Tight'' settings have been used for the basis set (for zinc the first hydrogen-like basis function of the second tier was additionally included), numerical integration grids and spatial basis function cutoff, with the self-consistency convergence criteria $10^{-6}$ for the charge density, $10^{-4}\,\mathrm{eV}$ for the eigenvalue sum, $10^{-8}\,\mathrm{eV}$ for the total energy and $10^{-6}\,\mathrm{eV\,\text{\AA{}}^{-1}}$ for the atomic forces.
As FHI-aims does not offer analytic Hessians, we use finite differences (FD) employing the tool ``get\_vibrations.py'' provided in the FHI-aims package. This tool generates displaced structures for the calculation of the FD Hessian. For each atomic Cartesian coordinate two displaced structures are computed with the displacements $+0.0025$ and $-0.0025\,\text{\AA{}}$, which are then used to construct the Hessian matrix. 

\section{Results}\label{sec:results}

\subsection{Model Systems}\label{sec:modelsystems}

\subsubsection{Structures}

\begin{figure}[!ht]
    \centering
    \includegraphics[scale=1.0]{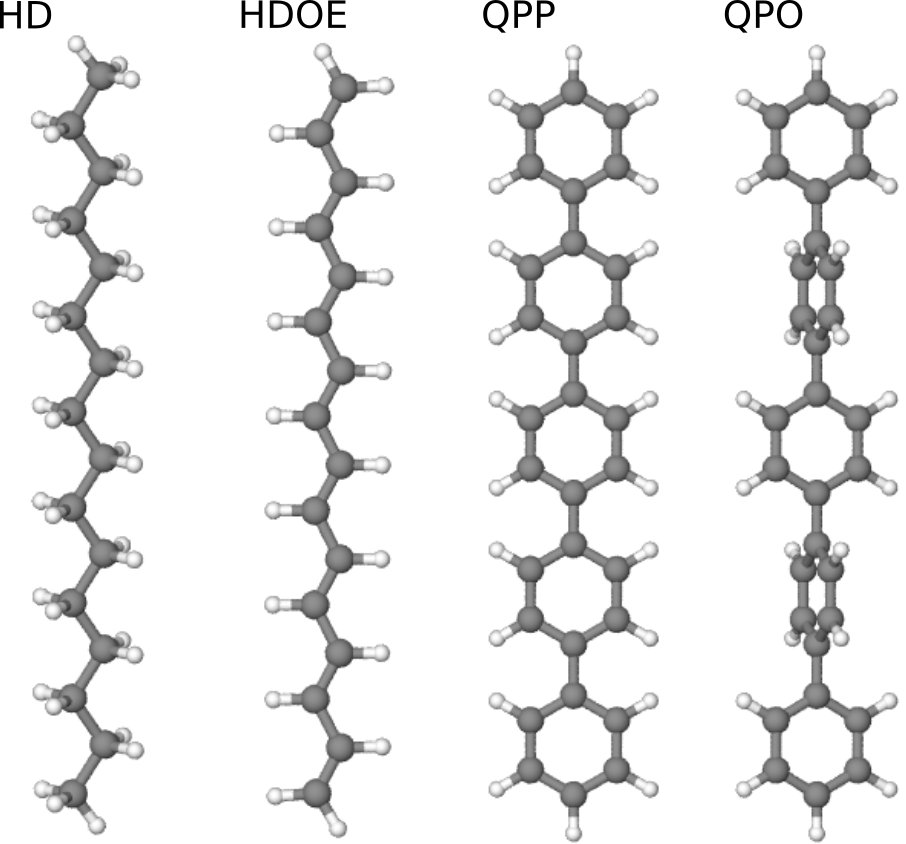}
    \caption{Model systems with different types of bonding used to investigate the atomic interactions. Hexadecane (HD) contains covalent single bonds only, while (3E,5E,7E,9E,11E,13E)-hexadeca-1,3,5,7,9,11,13,15-octaene (HDOE) represents a system with conjugated $\pi$-bonds. Further, two conformers of 1,1':4',1'':4'',1''':4''',1'''':4'''',1'''''-quinquephenyl are used, an all-in-plane conformer with maximum resonance stabilization of the $\pi$-electrons (QPP) and a conformer with alternating orthogonal phenylene rings (QPO) decoupling the aromatic subsystems.}
    \label{fig:Model_System_Structures}
\end{figure}

The range of the atomic interactions depends on the electronic structure of the system. Thus, we first investigate prototypical model systems covering various forms of covalent bonds including delocalized $\pi$-electrons in conjugated polyenes and aromatic molecules. These systems can be considered as idealized models representing different parts of more complex systems like MOFs and thus allow to study the effect of the electronic structure. Specifically, we have chosen the quasi one-dimensional molecules shown in Fig.~\ref{fig:Model_System_Structures} to investigate the distance dependence of the atomic interactions employing the Hessian. These model systems (Fig.~\ref{fig:Model_System_Structures}) include hexadecane (HD) as a typical molecule with single covalent bonds, (3E,5E,7E,9E,11E,13E)-hexadeca-1,3,5,7,9,11,13,15-octaene (HDOE) containing a conjugated $\pi$-electron system extending over the entire molecule and two conformers of 1,1':4',1'':4'',1''':4''',1'''':4'''',1'''''-quinquephenyl, an all-in-plane conformer (QPP) with maximum resonance stabilization of the $\pi$-electrons and a conformer with pairwise orthogonal phenylene rings (QPO), which prevents electronic resonance across the individual  subsystems due to the non-bonding overlap of the $p$-orbitals in neighboring rings. In case of HD and HDOE the structures have been fully optimized, while the QPP and QPO fragments have been derived from the structure of a relaxed benzene molecule, which has been replicated and connected using carbon-carbon bond lengths between rings of $1.45\,$\AA{}.

\subsubsection{Atomic Hessian Submatrix Norm}

\begin{figure*}[!hbt]
    \centering
    \includegraphics[scale=1.0]{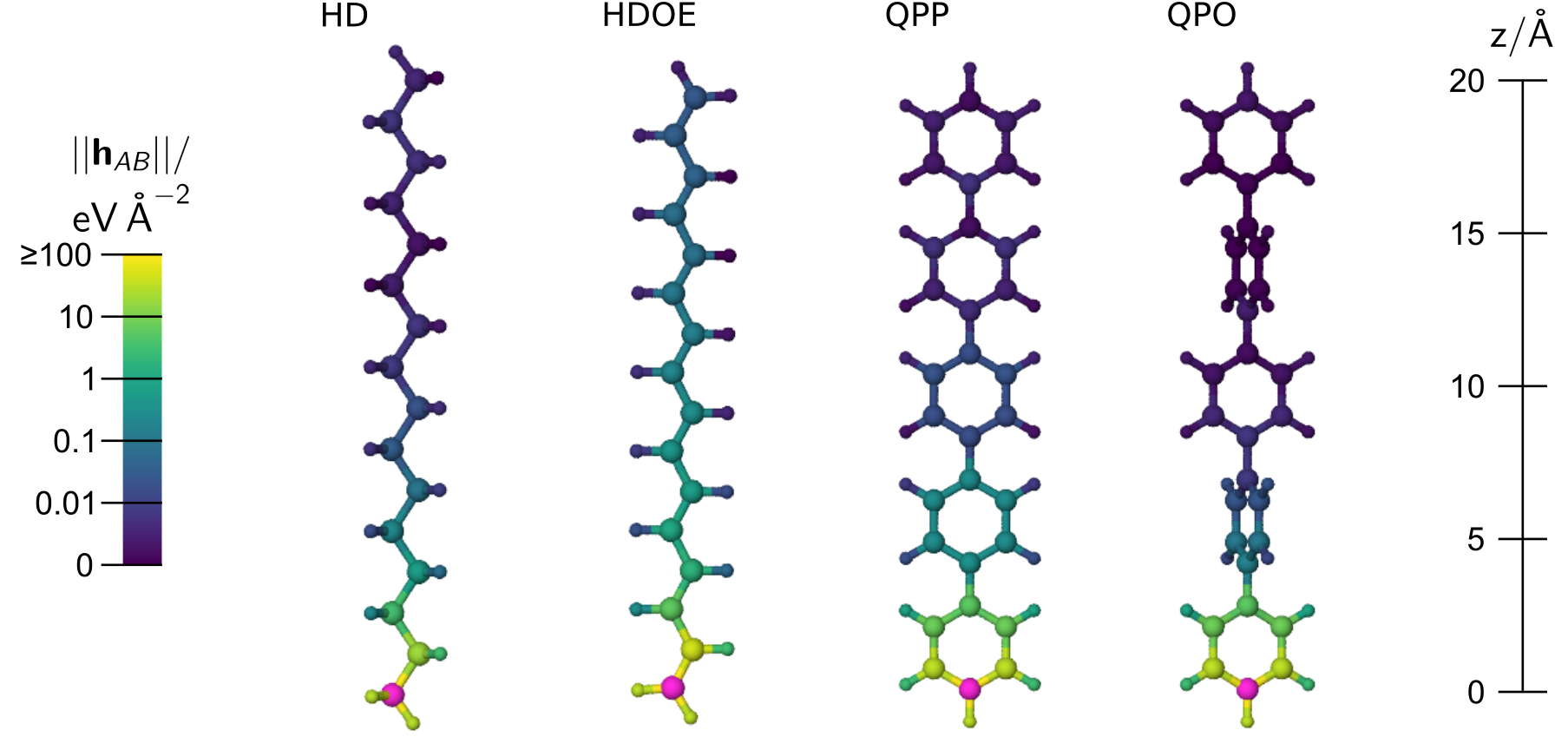}
    \caption{Atomic Hessian submatrix norm values $||\mathbf{h}_{AB}||$ describing the interaction between the magenta reference carbon atoms $A$ and all other atoms $B$ in the model systems HD, HDOE, QPP and QPO (logarithmic scale).}
    \label{fig:Model_System_HessianSubMatrixNormLOG}
\end{figure*}

\begin{figure}[ht!]
    \centering
    \includegraphics[width=\linewidth]{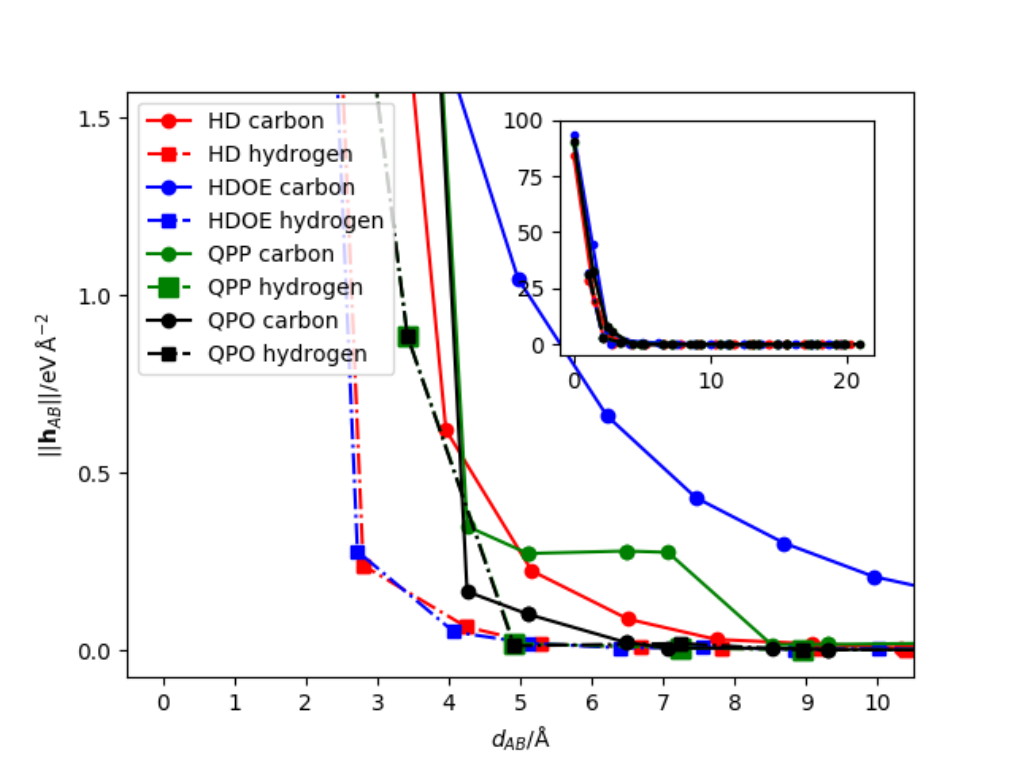}
    \caption{Atomic Hessian submatrix norm values $||\mathbf{h}_{AB}||$ of the four model systems HD, HDOE, QPP and QPO as a function of the distance $d_{AB}$ between the reference carbon atom $A$ as defined in Fig. \ref{fig:Model_System_HessianSubMatrixNormLOG} and all other atoms $B$. Separate curves are given for the interactions of atom $A$ with neighboring carbon and hydrogen atoms. The inset shows the data for the interaction of $A$ with all atoms in the entire molecules.}
    \label{fig:Model_System_distanceVSnorm_ALL}
\end{figure}

As a first step, we have investigated the range of the atomic interactions in these model systems by analyzing the Hessian obtained from DFT.
The values of the atomic Hessian submatrix norm of all atoms with respect to the corresponding terminal reference carbon atoms $A$ highlighted in magenta are shown in Fig.~\ref{fig:Model_System_HessianSubMatrixNormLOG} using a logarithmic scale. As expected, the atomic Hessian submatrix norm decreases with increasing interatomic distance, reflecting the decaying influence of the neighboring atoms on the force $\mathbf{f}_A$ acting on the reference atoms.  Fig.~S1
in the electronic SI shows the same data using a linear scale. Moreover, Figs.~\ref{fig:Model_System_distanceVSnorm_ALL} and S2
show the decay of the atomic Hessian submatrix norm $||\mathbf{h}_\mathrm{AB}||$ with increasing atomic distance $d_\mathrm{AB}$ for the neighboring hydrogen and carbon atoms. Of course, apart from the coordinates of the other atoms in the systems, the position of the reference atom itself has a strong influence on the force acting on it, which is described by the on-site submatrices $\mathbf{h}_{AA}$.

Although the qualitative results are similar for all investigated model systems, the electronic structure plays an important role for the quantitative decay of the interactions, since in particular for the carbon atoms the decay for HDOE and QPP is much slower, i.e., the interaction at larger distances is stronger, compared to HD and QPO. On the other hand, for a given distance the reference carbon atoms in all model systems show smaller interactions with the hydrogen atoms compared to the carbon atoms. This can be understood by the larger number of bonds mediating the interactions of the hydrogen atoms with the carbon reference atoms. More importantly, the hydrogen atoms do not participate in $\pi$-systems, which are particularly relevant for significant interactions over larger distances. Consequently, in Fig.~\ref{fig:Model_System_distanceVSnorm_ALL} the hydrogen curves of HD and HDOE are quite similar to each other and also the hydrogen curves of the QPP and QPO molecules are almost indistinguishable, but larger in magnitude compared to the alkane and alkene systems indicating a still more efficient mediation of the hydrogen interactions within the first aromatic ring.  

For the carbon atoms, the effect of the $\pi$-system is very pronounced, and as a consequence the interactions with the reference atom are stronger in HDOE compared to HD. Furthermore, the electron delocalization, which is increased in QPP compared to QPO due to the in-plane conformation of the phenylene rings, affects the interactions with the reference atom. The QPP and QPO carbon curves start to differ at distances beyond about $4\,$\AA{}, since the structure of the first phenyl ring is identical in both systems. For the carbon atoms of the second phenylene ring, a plateau in the QPP curve is observed, which indicates a similar iteraction with all these atoms and which is not present in the QPO curve, since in the latter system the interactions are efficiently truncated by the orthogonality of the aromatic subsystems.

\subsubsection{Hessian Group Matrix Norm}

\begin{figure*}[!ht]
    \centering
    \includegraphics[width=\linewidth]{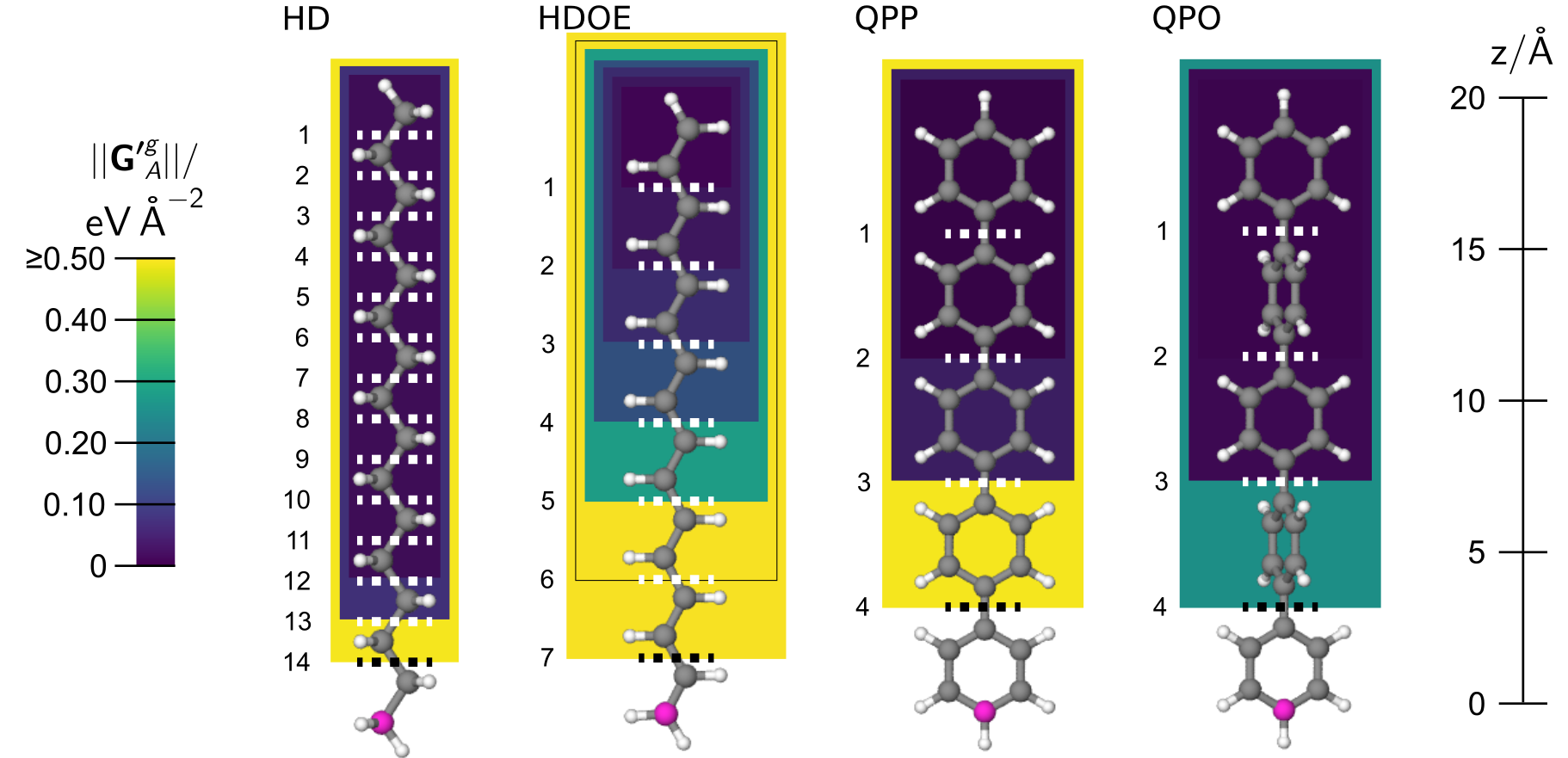}
    \caption{Effective Hessian group matrix norm $||\mathbf{G'}_{A}^g||$ for all atomic groups (represented by the colored rectangles) with respect to the reference carbon atoms $A$ shown in magenta for the model systems HD, HDOE, QPP and QPO. The bonds, which are cut to form increasing groups of removed atoms, are shown as white dashed lines along with the numbering of the resulting groups from the top to the bottom.  The black dashed lines indicate the bond to be cut for the smallest considered fragment corresponding to the largest atomic group of removed atoms. Note that each group is included in the next larger group when more atoms are removed from the system.}
    \label{fig:Model_Systems_HessianGroupExSubMatrixNorm}
\end{figure*}

The atomic Hessian submatrix norm shows a decreasing interaction of the atoms with increasing interatomic distance. In principle this could allow us to define a threshold value to decide, which atoms are only weakly interacting and thus can be eliminated from the fragment without a substantial loss in accuracy of $\mathbf{f}_A^{\mathrm{frag}}$ obtained in the DFT calculations. However, strictly applying such a threshold would result in the cutting of arbitrary bonds, which can give rise to significant changes in the electronic structure, e.g. if $\pi$-bonds are broken. These changes in the electronic structure can severely affect the atomic interactions, i.e., the energy and forces of the system. Consequently, the atomic Hessian submatrix norm cannot be used directly as a criterion for arbitrary systems. Instead, chemically meaningful functional groups have to be defined, which must not be cut but have to be included or excluded entirely. If these functional groups are considered in the construction of the molecular fragments, the joint effect of the eliminated atoms can be described to a good approximation by the effective Hessian group matrix norm.

With this procedure, the electronic structure of the model systems remains as intact as possible, and the resulting functional groups can be removed step by step from the large reference systems shown in Fig.~\ref{fig:Model_Systems_HessianGroupExSubMatrixNorm}. While for the simple alkane HD each CH$_2$ group can be individually removed, the conjugated $\pi$-bonds in HDOE require the removal of C$_2$H$_2$ entities. For the aromatic systems, each phenylene ring forms a functional group that can be eliminated. Moreover, we avoid cutting bonds involving the reference atoms, which results in the smallest possible fragment ethane for the HD system. Hence, there are 15, 8, 5 and 5 fragments, respectively, that can be built for the four model systems. For this purpose, the functional groups are cut sequentially to form fragments of decreasing size, while the number of removed atoms increases, resulting in 14, 7, 4 and 4 different groups of removed atoms. We note that each group of eliminated atoms is included in the next larger group such that according to the numbering scheme in Fig.~\ref{fig:Model_Systems_HessianGroupExSubMatrixNorm} a group with larger number contains all atoms in the groups with a smaller number. 

The colors of the boxes containing all atoms of a given group in Fig.~\ref{fig:Model_Systems_HessianGroupExSubMatrixNorm} represent the values of the effective Hessian group matrix norm computed according to Eq.~\ref{eq:hessian_extgroup_sub_matrix}. Similar to the atomic Hessian submatrix norm discussed above, the effective Hessian group matrix norm decreases with increasing distance of the group of atoms from the reference atom $A$ for all model systems. Its value can now be related to the error of the force 
\begin{equation}
    \Delta \mathbf{f}^{Y_g}_A = \mathbf{f}^{Y}_A - \mathbf{f}^{\mathrm{frag,}Y_g}_A
\end{equation}
for system $Y=\{$HD,HDOE,QPP,QPO$\}$, which is the difference of the force $\mathbf{f}_A^Y$ in the full system and the force $\mathbf{f}_A^{\mathrm{frag,}Y_g}$ when removing group $g$ and saturating the fragment by hydrogen, which is placed along the broken carbon-carbon bond with a carbon-hydrogen distance of $1.05\,$\AA{}. 
Fig.~S3
in the SI shows the force error and the effective Hessian group matrix norm for each fragment size for the four model systems providing clear evidence for a similar behavior of both quantities. 

Finally, we have investigated the effect of the saturating hydrogen atom on the Hessian group matrix norm taking the HD system as example (Fig.~S4). We found that for distances larger than approximately 5~\AA{} its contribution to the effective Hessian group matrix norm is very small, such that data sets typically used for the construction of machine learning potentials employing common environment radii of 5-6 \AA{} will only be marginally affected by the hydrogen saturation.

\subsubsection{Force Convergence Threshold}

\begin{figure*}[!ht]
    \centering
    \includegraphics[width=\linewidth]{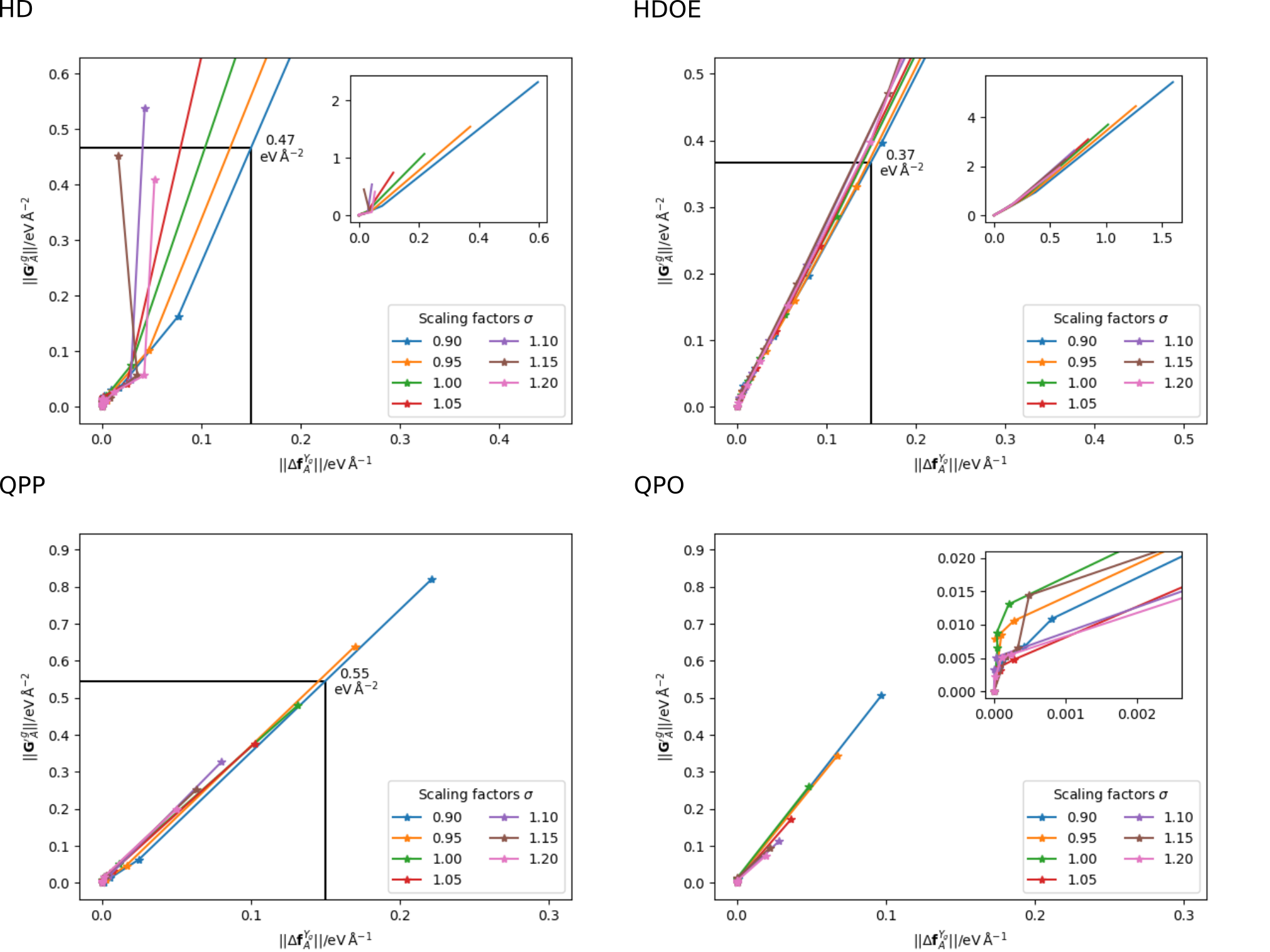}
    \caption{Effective Hessian group matrix norm $||\mathbf{G'}_{A}^g||$ as a function of the norm of the force error $||\Delta \mathbf{f}^{Y_g}_A||$ of the reference carbon atom $A$ in the four model systems HD, HDOE, QPP and QPO. For HD and HDOE the inset shows the complete data range, while for QPO the enlarged data near the origin is shown.
    For each system the data is shown for different structural scaling factors $\sigma =0.90-1.20$.  For each $\sigma$ the smallest fragment belongs to the largest $||\mathbf{G'}_{A}^g||$, i.e., the fragment size increases towards the origin corresponding to a decreasing force error with respect to the full system. The black lines in the panels of HD, HDOE and QPP show the largest $||\mathbf{G'}_{A}^g||$ value compatible with the employed force convergence criterion $||\Delta \mathbf{f}^\mathrm{max}||=0.15\,\mathrm{eV\,\text{\AA{}}^{-1}}$, which in case of the QPO system is met by all fragments.}
    \label{fig:Model_Systems_relation-hessian-force}
\end{figure*}

The correlation between the force error $||\Delta \mathbf{f}^{Y_g}_A||$ and the effective Hessian group matrix norm $||\mathbf{G'}_{A}^g||$ is shown in Fig.~\ref{fig:Model_Systems_relation-hessian-force} for the different fragments of all model systems. Since, as described above, all these structures are close to equilibrium geometries, in order to obtain a more diverse data set, we have also included non-equilibrium structures generated by scaling the geometries of all four model systems employing scaling factors $\sigma = 0.90 - 1.20$. The data for these compressed and expanded molecules is shown along with the data of the relaxed model systems ($\sigma=1.00$) in Fig.~\ref{fig:Model_Systems_relation-hessian-force}.

There is an approximately linear relationship, and even when including strongly distorted molecular structures the close-to linear correlation between both quantities holds.  The most prominent deviations from linearity are found for the smallest expanded HD fragments, i.e., ethane (HD in Fig.~\ref{fig:Model_Systems_relation-hessian-force}), which is reasonable as in this case already the second-nearest carbon atoms of the reference atoms are removed. Still, overall the force errors are surprisingly small even in this situation. The largest force errors are found for the most compressed molecules (blue curves), in line with large absolute forces resulting from the strong atomic repulsion.

By defining the desired convergence level of the forces, a threshold $\Gamma$ for the effective Hessian group matrix norm can now be derived to distinguish between important atoms that should be included in the fragments and weakly interacting atoms that can be omitted. However, due to the different types of bonding in the investigated model systems it is not yet clear to what extent such a threshold will depend on the specific system. In case of a pronounced system-dependence such a threshold would be of little use, since for each system and maybe even each atomic environment a different threshold would have to be determined. If, however, a similar threshold applicable to all model systems could be identified, its value could be employed to determine the minimum fragment size for a wide range of systems with very different types of bonding.

We now choose a convergence criterion for the force error of $||\Delta \mathbf{f}^{Y_g}_A||=0.15\,$eV\,\AA{}$^{-1}$, which is the typical accuracy that can be achieved for the root mean squared error of forces in MLPs. Using this criterion, we can determine the corresponding value of $\Gamma$ (see black lines in Fig.~\ref{fig:Model_Systems_relation-hessian-force}) for the systems with the largest force errors, i.e., the most compressed molecules ($\sigma=0.9$ for HD, HDOE and QPP). We note that for the QPO system, even for the smallest fragment, i.e., benzene, the force error is below the targeted convergence level since the reference carbon atom is only very weakly interacting with the neighboring orthogonal phenylene ring of the system.

From the data in Fig.~\ref{fig:Model_Systems_relation-hessian-force} we find surprisingly similar $\Gamma$ values for the effective Hessian group matrix norm (HDOE: $0.37\,\mathrm{eV\,\text{\AA{}}^{-2}}$, HD: $0.47\,\mathrm{eV\,\text{\AA{}}^{-2}}$, QPP: $0.55\,\mathrm{eV\,\text{\AA{}}^{-2}}$) for the different bonding situation in the model systems. This implies that -- to a good approximation -- a general threshold value can be defined that is applicable to a wide range of systems. Based on the tightest criterion $0.37\,\mathrm{eV\,\text{\AA{}}^{-2}}$, i.e., the force convergence of the HDOE model system exhibiting the strongest long-range interactions, and taking into account the estimated accuracy of the effective Hessian group matrix norm values of $\pm0.02\,\mathrm{eV\,\text{\AA{}}^{-2}}$ as a safety margin, we now define a threshold value of $\Gamma=0.35\,\mathrm{eV\,\text{\AA{}}^{-2}}$. Below this value also the force vectors in all other systems are well converged to the desired accuracy.

In the next step we now use this $\Gamma$ value to identify the minimum fragment sizes in all model systems required to reach converged force vectors of the reference carbon atoms, which are the fragments HDOE$_{5}$ ((1E,3E,5E)-hexatriene) (Tab.~\ref{tab:SC_HDOE}), HD$_{13}$ (propane) and QPP$_{3}$ (biphenyl) (Tab.~S1),
respectively. In case of QPO (Tab.~S1) all fragments fulfill the convergence criterion such that even the smallest fragment, i.e., QPO$_4$ (benzene), is converged.

\begin{table}[!htb]
     \centering
     \begin{ruledtabular}
     \caption{
Compilation of the force component errors $\Delta f^{\mathrm{HDOE}_g}_{A_{x,y,z}}$ and the total force errors $||\Delta \mathbf{f}^{\mathrm{HDOE}_g}_{A}||$ in $\mathrm{eV\,\text{\AA{}}^{-1}}$ for the reference carbon atom in the model system $Y=\mathrm{HDOE}$ (Fig.~\ref{fig:Model_Systems_HessianGroupExSubMatrixNorm} and \ref{fig:Model_Systems_relation-hessian-force}, $\sigma=1.00$). Further, the effective Hessian group matrix norm $||\mathbf{G'}_{A}^{g}||$ is given in $\mathrm{eV\,\text{\AA{}}^{-2}}$. Numbers outside the intended convergence criterion are given in bold.}
     \label{tab:SC_HDOE}
     \begin{tabular}{cccccc}
        $g$	&	$\Delta f^{\mathrm{HDOE}}_{A_\mathrm{x}}$	&	$\Delta f^{\mathrm{HDOE}_g}_{A_\mathrm{y}}$	&	$\Delta f^{\mathrm{HDOE}_g}_{A_\mathrm{z}}$ & $||\Delta\mathbf{f}^{\mathrm{HDOE}_g}_{A}||$	    &	$||\mathbf{G}'^{g}_{A}||$	\\
\hline
ref	&	0.0000	&	0.0000	&	$0.0000$	&	0.0000	&	0.00							\\
1	&	0.0000	&	0.0014	&	$-0.0043$	&	0.0045	&	0.02							\\
2	&	0.0000	&	0.0042	&	$-0.0116$	&	0.0124	&	0.04							\\
3	&	0.0000	&	0.0094	&	$-0.0247$	&	0.0264	&	0.07							\\
4	&	0.0000	&	0.0195	&	$-0.0495$	&	0.0532	&	0.14							\\
5	&	0.0000	&	0.0414	&	$-0.1031$ &	0.1111	&	0.29				\\
6	&	0.0000	&	0.0986	&	$\mathbf{-0.2474}$	&	$\mathbf{0.2663}$	&	$\mathbf{0.71}$	\\
7	&	0.0000	&	$\mathbf{0.1877}$	&	$\mathbf{-1.0018}$	&	$\mathbf{1.0192}$	&	$\mathbf{3.71}$	\\
     \end{tabular}
     \end{ruledtabular}
 \end{table}

Hence, we obtain significantly smaller fragments required for the HD and QPO systems compared to HDOE and QPP. In order to obtain forces converged to within 0.15~$\mathrm{eV\,\text{\AA{}}^{-1}}$, fragments of a radius of approximately 2.6, 6.2\textit{}, 7.1, and 2.8~\AA{} are found necessary for HD, HDOE, QPP and QPO, respectively.

\subsection{MOF-5}\label{sec:DFTMOF5}

\begin{figure*}[!hp]
    \centering
    \includegraphics[scale=1.0]{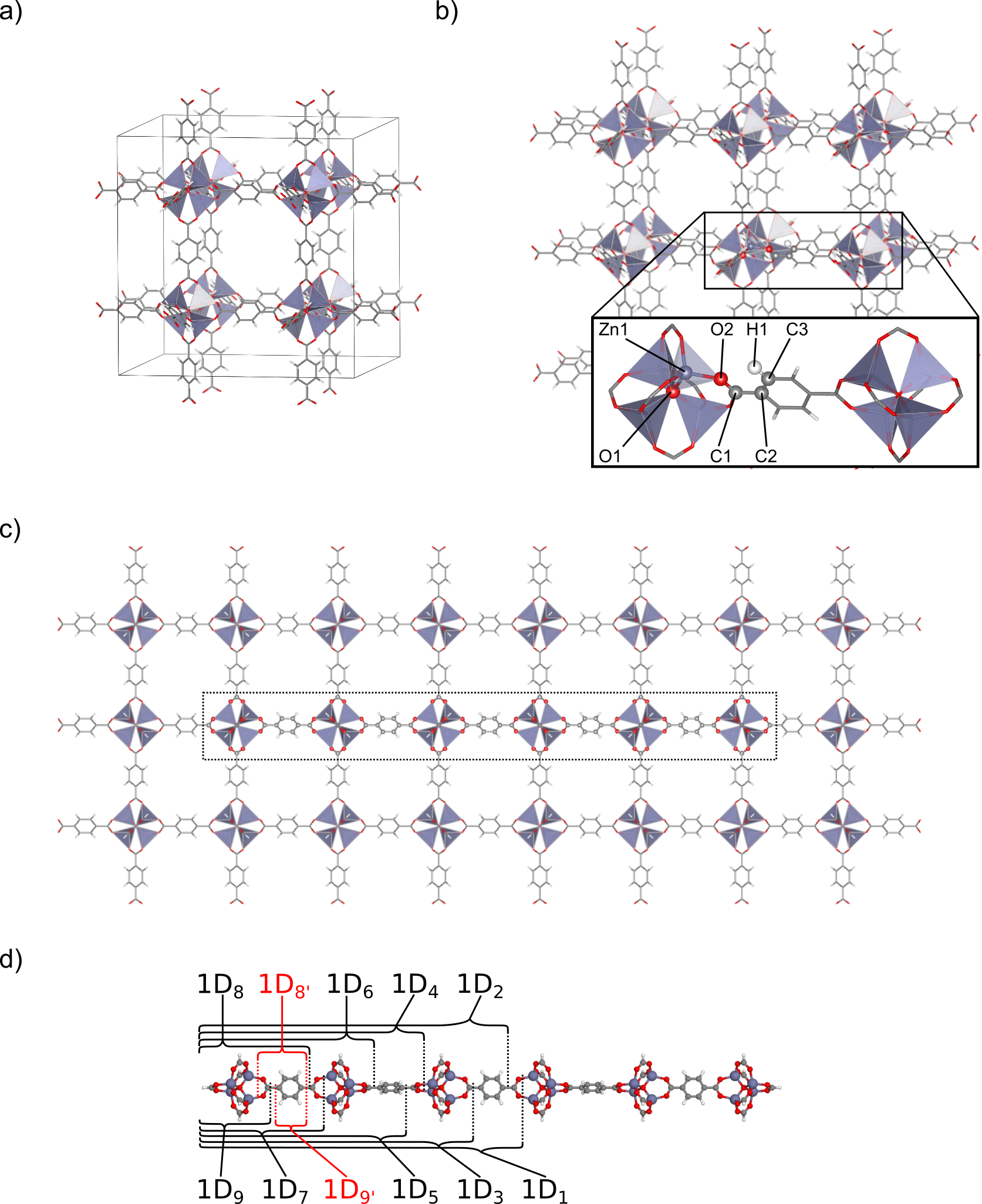}
    \caption{a) Crystal structure of MOF-5 containing eight Zn$_4$O(BDC)$_3$ formula units (BDC = benzene-1,4-dicarboxylate) per unit cell. Panel b) shows the inequivalent atoms in the bulk crystal as spheres including the atom labeling used in the present work, whereas the remaining structure is represented by sticks. In c) a one-dimensional model system (``1D'') consisting of six SBUs and five linker molecules in the periodic MOF-5 crystal is highlighted. The extracted and hydrogen-saturated 1D model system is shown in d) and several fragments of different size labeled as 1D$_\mathrm{1}$, $\hdots$, 1D$_\mathrm{9}$ are constructed. Since the smallest of these fragments do not contain reference atoms in the linker molecule, additional fragments 1D$_\mathrm{8'}$ and 1D$_\mathrm{9'}$ are defined for this case. Atomic colors: Zn violet, O red, C gray and H white.}
    \label{fig:mof5}
\end{figure*}

Having identified a threshold value for the norm of the effective Hessian group matrix ensuring converged forces for different model systems, we now turn to the more complex metal-organic-framework MOF-5 to assess the general applicability of these thresholds. The cubic unit cell of this material with the space group $Fm\overline{3}m$ (space group no. 225) is shown in Fig.~\ref{fig:mof5}a. It consists of eight secondary building units (SBUs, $\mathrm{Zn_4O}$) and 24 connecting linker molecules (BDC = benzene-1,4-dicarboxylate), i.e., eight formula units of Zn$_\text{4}$O(BDC)$_\text{3}$. Due to the symmetry of the crystal, bulk MOF-5 contains seven inequivalent atomic positions labeled Zn1, O1, O2, C1, C2, C3 and H1 as shown in Fig.~\ref{fig:mof5}b.

\subsubsection{One-dimensional MOF-5 Fragments}\label{sec:1D-DFTMOF5}

Before investigating the three-dimensional MOF-5 structure, we start with a one-dimensional model system (``1D'') cut from the bulk consisting of six SBUs connected by five linker molecules and satured by hydrogen (Fig.~\ref{fig:mof5}c). The fragments $\mathrm{1D}_g$ including reference atoms located in the SBU are labeled 1D$_\mathrm{1}$, ..., 1D$_\mathrm{9}$ in the order of decreasing fragment size or, equivalently, increasing number of atoms in the group of removed atoms (Fig. \ref{fig:mof5}d). Moreover, two fragments 1D$_\mathrm{8'}$ and 1D$_\mathrm{9'}$ are constructed for reference atoms in the phenylene rings when applying very small environment radii. A more detailed description of the fragment construction and the structural entities that have to be preserved in this system is given in Sec.~S-II in the supporting information.

\begin{figure*}[!ht]
    \centering
    \includegraphics{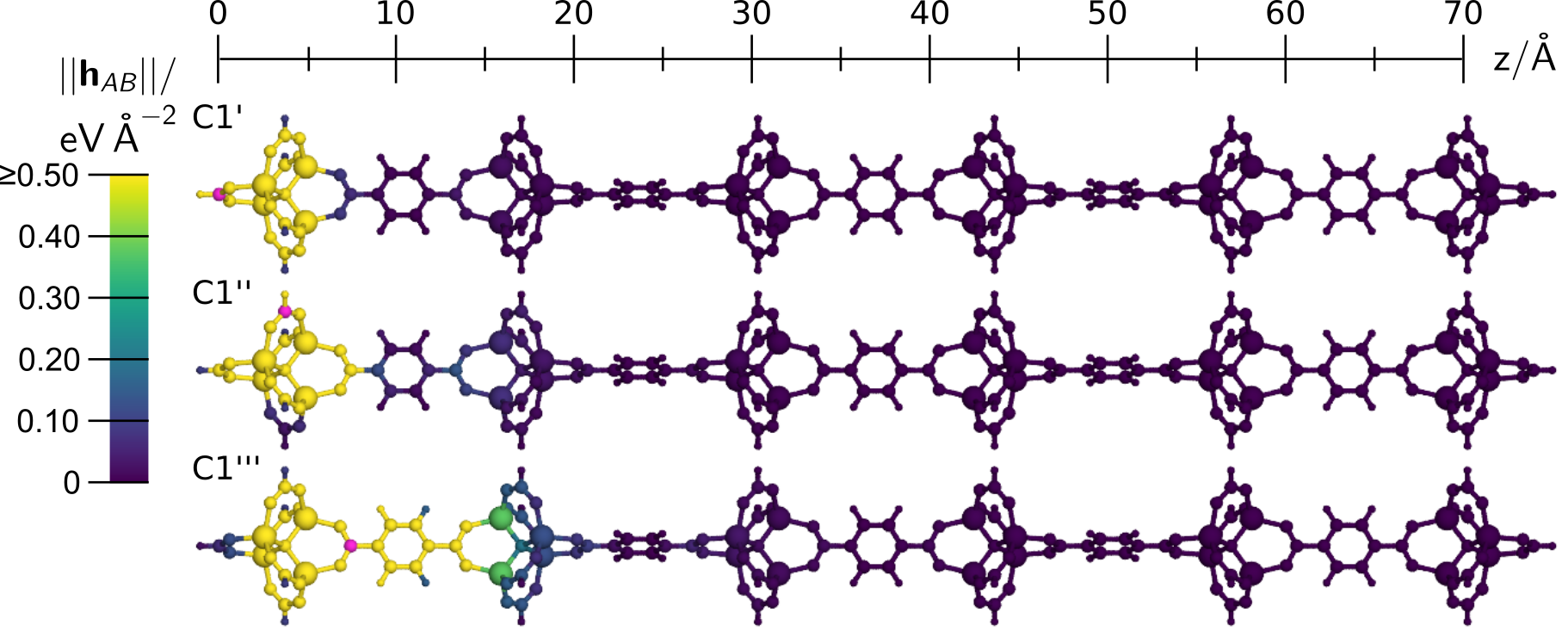}
    \caption{Atomic Hessian submatrix norm values $||\mathbf{h}_{AB}||$ for three different reference carbon atoms $\mathrm{C1'}$, $\mathrm{C1''}$, and $\mathrm{C1'''}$ (magenta color) corresponding to atom C1 in bulk MOF-5.}
    \label{fig:1D-MOF_Hessian_atomic_norms-C1}
\end{figure*}

\begin{figure*}[!ht]
    \centering
    \includegraphics{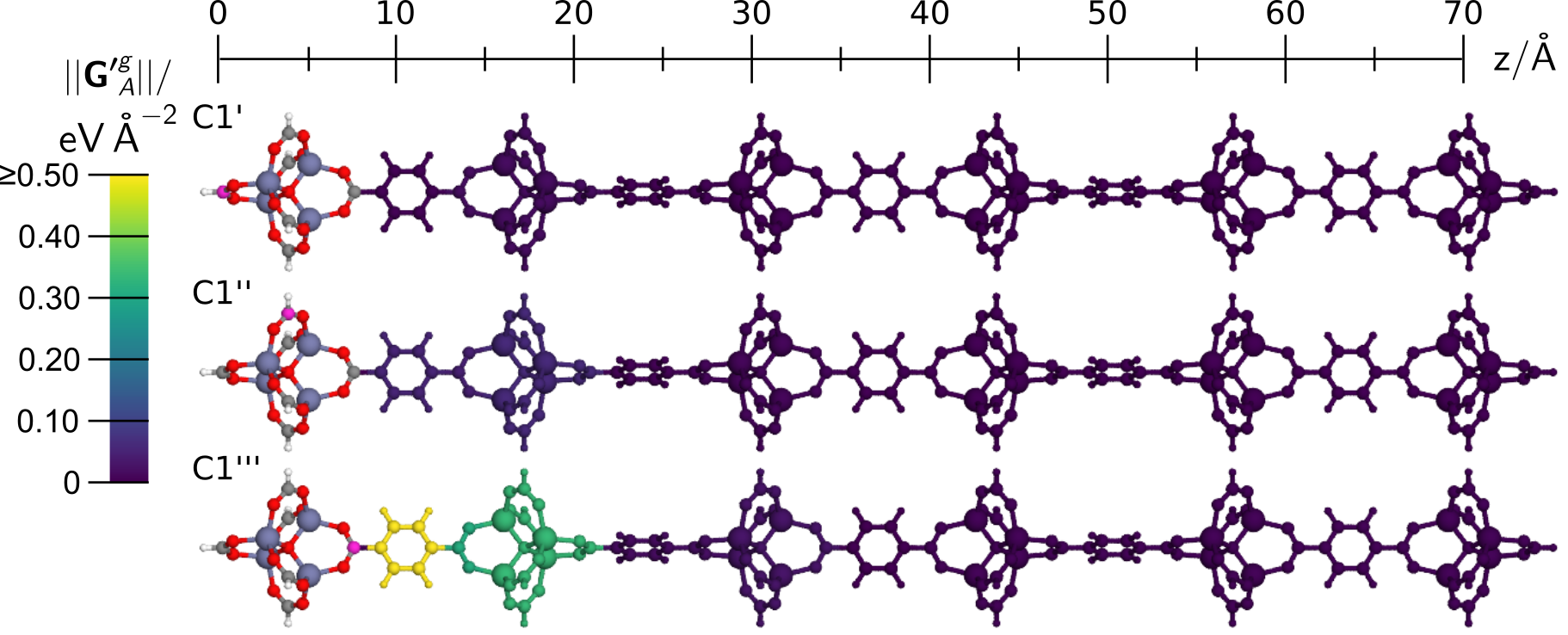}
    \caption{Effective Hessian group matrix norm $||\mathbf{G'}^g_{A}||$ for three different reference carbon atoms $\mathrm{C1'}$, $\mathrm{C1''}$, and $\mathrm{C1'''}$ (magenta color, see also Fig.~\ref{fig:1D-MOF_Hessian_atomic_norms-C1}) corresponding to atom C1 in bulk MOF-5. The color of all atoms of a group is defined by $||\mathbf{G'}^g_{A}||$ with $g$ as the smallest sub group the atom belongs to. The atomic colors of the smallest possible fragment, i.e., the SBU at the very left, are specified by the atom's element.}
    \label{fig:1D-MOF_Hessian_extGROUP_norms-C1}
\end{figure*}

For the example of atomic site C1 the analysis of the atomic Hessian submatrix norm is shown in the three panels of Fig.~\ref{fig:1D-MOF_Hessian_atomic_norms-C1} corresponding to three possible choices $\mathrm{C1}'$, $\mathrm{C1}''$ and $\mathrm{C1}'''$ resulting from the reduced symmetry of the one-dimensional fragment. Similar analyses for the remaining atomic sites are shown in Fig.~S5 and S6
in the SI. We find that in all cases atoms close to the reference atom, e.g. within the same SBU, exhibit relevant interactions. Further, in particular reference atoms included in or close to the $\pi$-system of the linker are interacting with their environment over rather long distances. This is consistent with the properties of, e.g., the aromatic QPP model system reported above. On the other hand, the SBU with stronger ionic bonding contributions is to some extent screening the atomic interactions resulting in a reduced range (see also Fig.~S5 and S6).
The same phenomenon is observed in the analysis of the effective Hessian group matrix norm in Fig.~\ref{fig:1D-MOF_Hessian_extGROUP_norms-C1} showing a significant interaction with the linker and the next nearest SBU only for $\mathrm{C1'''}$ in the bridging carboxylic group that is already participating in the $\pi$-system (see also Fig.~S7 and S8
 for the other atomic sites).
 
\begin{table}[!ht]
     \centering
     \caption{Compilation of the force component errors $\Delta f^{\mathrm{1D}_g}_{\mathrm{C1'''}_{x,y,z}}$ and the total force errors $||\Delta \mathbf{f}^{\mathrm{1D}_g}_{\mathrm{C1'''}}||$ in $\mathrm{eV\,\text{\AA{}}^{-1}}$ for the C1$'''$ reference atom in different fragments of the one-dimensional MOF-5 system 1D shown in Fig. \ref{fig:1D-MOF_Hessian_atomic_norms-C1}. Further, the effective Hessian group matrix norm $||\mathbf{G'}_{\mathrm{C1'''}}^{g}||$ is given in $\mathrm{eV\,\text{\AA{}}^{-2}}$. Numbers outside the intended convergence are given in bold.}
     \label{tab:SC_1D_Structure_C1}
     \begin{ruledtabular}
     \begin{tabular}{ccccccc}
        $g$	&	$\Delta f^{\mathrm{1D}_g}_{\mathrm{C1'''}_{x}}$	&	$\Delta f^{\mathrm{1D}_g}_{\mathrm{C1'''}_{y}}$	&	$\Delta f^{\mathrm{1D}_g}_{\mathrm{C1'''}_{z}}$	    & $||\Delta\mathbf{f}^{\mathrm{1D}_g}_{\mathrm{C1'''}}||$	    &	$||\mathbf{G}'^{g}_{\mathrm{C1'''}}||$	\\
\hline
ref	&	0.0000	&	0.0000	&	0.0000	&	0.0000	&	0.00							\\
1	&	0.0001	&	$-0.0001$	&	0.0000	&	0.0002	&	0.01							\\
2	&	0.0000	&	0.0000	&	$-0.0013$	&	0.0013	&	0.01							\\
3	&	0.0000	&	0.0000	&	0.0001	&	0.0001	&	0.00							\\
4	&	0.0001	&	0.0001	&	0.0005	&	0.0005	&	0.03							\\
5	&	0.0000	&	0.0000	&	$-0.0069$	&	0.0069	&	0.02							\\
6	&	0.0000	&	0.0000	&	0.0016	&	0.0016	&	0.02							\\
7	&	0.0006	&	$-0.0006$	&	0.0003	&	0.0008	&	0.34							\\
8	&	0.0000	&	0.0000	&	$-0.1492$	&	0.1492	&	0.31				\\
9	&	$-0.0003$	&	0.0001	&	$\mathbf{3.0328}$	&	$\mathbf{3.0328}$	&	$\mathbf{21.80}$\\
     \end{tabular}
     \end{ruledtabular}
\end{table}

Next, we investigate if the threshold $\Gamma$ derived from the model systems can be used to identify size-converged fragments in the 1D system. Table \ref{tab:SC_1D_Structure_C1} shows the fragment results for reference atom $\mathrm{C1'''}$ (Fig. \ref{fig:1D-MOF_Hessian_extGROUP_norms-C1}). Applying the previously determined threshold $\Gamma=0.35\,\mathrm{eV\,\text{\AA{}}^{-2}}$ predicts a fragment of size 1D$_8$ or larger to be required to obtain a converged total force, which is confirmed by the corresponding force error of $||\Delta\mathbf{f}^{\mathrm{1D}_8}_{\mathrm{C1'''}}||=0.1492~\mathrm{eV\,\text{\AA{}}^{-1}}$ that is indeed below the threshold of $0.15~\mathrm{eV\,\text{\AA{}}^{-1}}$. In conclusion, the threshold determined for the four model systems is found to be applicable also to one-dimensional fragments of MOF-5 ensuring accurately converged force vector components.

\subsubsection{Three-Dimensional MOF-5 Fragments}\label{sec:3D-DFTMOF5}

\begin{figure*}[!ht]
    \centering
    \includegraphics[width=\linewidth]{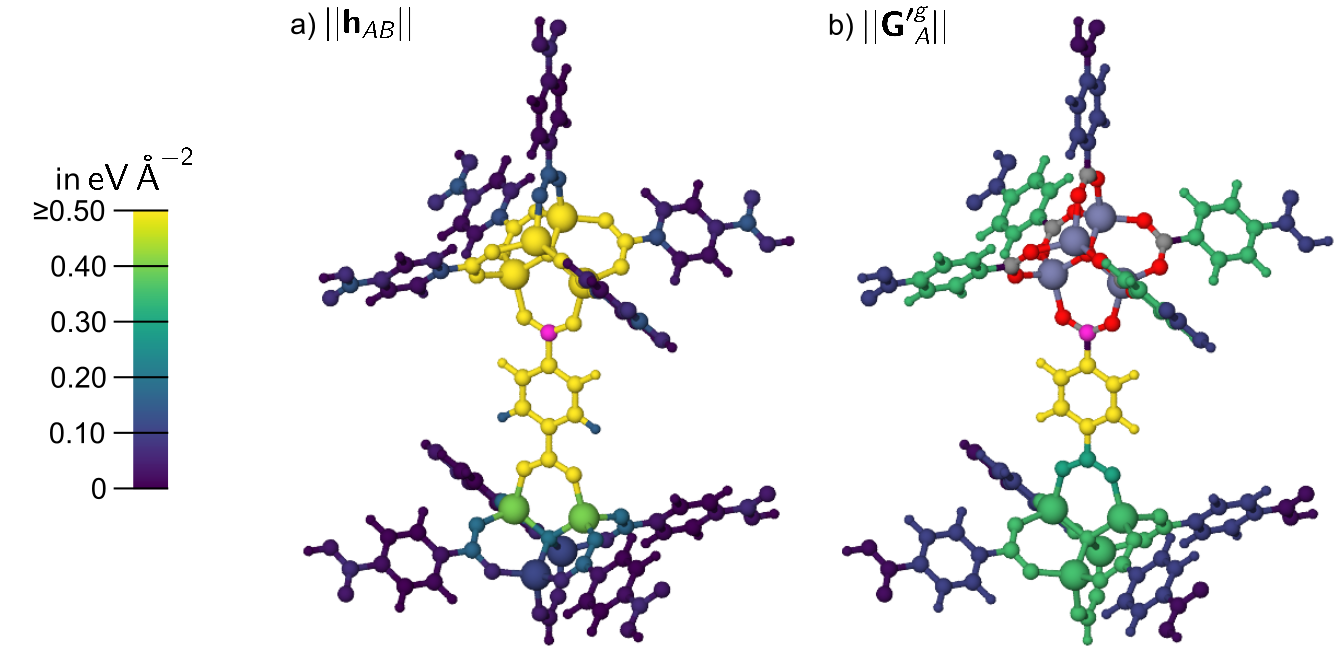}
    \caption{ a) Atomic Hessian submatrix norm values $||\mathbf{h}_{AB}||$ and b) effective Hessian group matrix norm values $||\mathbf{G'}_{A}^g||$ in $\text{eV\,\AA{}}^{-2}$ with respect to the central atom $A=\mathrm{C1}$ (magenta) in reference structure $\mathrm{C1_{ref}}$. $||\mathbf{G'}_{\mathrm{A}}^g||$ defines the color for the closest atoms of a given group, which in addition also contains all atoms at larger distance. The colors of the smallest possible fragment in b) refer to the chemical elements.}
    \label{fig:FDHessianC1}
\end{figure*}

Finally, we now address the three-dimensional structure of MOF-5. Instead of performing calculations of the periodic bulk system, which might be computationally unfeasible for even larger systems or might suffer from artificial periodicity in case of smaller systems, we construct very large molecular fragments centered at the seven inequivalent atomic sites (see also the discussion in S-V in the SI).
The required size of the reference fragments is determined using the atomic Hessian submatrix norm values. These should be close to zero for the outermost atoms with respect to the central atoms. In practice, from the 1D reference structure we found values $<0.1~\mathrm{eV\,\text{\AA{}}^{-2}}$ yielding well-converged reference fragments C1$_\mathrm{ref}$, Zn1$_\mathrm{ref}$, O1$_\mathrm{ref}$, O2$_\mathrm{ref}$, C2$_\mathrm{ref}$, C3$_\mathrm{ref}$, and H1$_\mathrm{ref}$ with $r_\mathrm{frag} = 10 - 12\,$\AA{} corresponding to the entire structures shown in Figs. \ref{fig:FDHessianC1}  and S10
to S15, which we use as starting points to construct smaller trial fragments.

As an example, we will now discuss the environment of site C1. Figure \ref{fig:FDHessianC1} shows the atomic Hessian submatrix norm of all atoms in C1$_\mathrm{ref}$ in panel a) and the effective Hessian group matrix norm in panel b). 
Like in case of the model systems and the 1D MOF-5 fragment, we find decreasing atomic Hessian submatrix norm values with increasing atomic distance, and also the effective Hessian group matrix norm decreases similarly to the case of the 1D fragment ($\mathrm{C1'''}$ in Fig. \ref{fig:1D-MOF_Hessian_extGROUP_norms-C1}). Due to the three-dimensional structure of the reference fragment, each group of atoms forms a shell-like structure around the reference atom, and each group also contains the atoms of the more distant groups.

We note that in the three-dimensional case each group contains a much larger number of atoms compared to the 1D system, which increases the values of the effective Hessian group matrix norm reflecting a potentially stronger overall influence of the environment on the force of the central atom that must be taken into account in the choice of the fragment size. The fragments C1$_\mathrm{1}$ -- C1$_\mathrm{7}$ used for the analysis of site C1 are shown in Fig.~S9,
while the corresponding errors of the forces and force components as well as the norm values of the effective Hessian group matrix are compiled in Table \ref{tab:SC_C1}. Using the threshold $\Gamma$ derived from the model systems, we find that fragment C1$_\mathrm{3}$ is size-converged. In view of the force errors, this is a rather safe estimate, since already the smaller fragment C1$_\mathrm{5}$ could be considered as converged based on the values in Table \ref{tab:SC_C1}. Still, as discussed in the introduction, the early convergence of forces might be artificial for particular (e.g. symmetric) environments, which is avoided when using Hessian-based quantities.

\begin{table}[!ht]
     \centering
     \caption{ Errors of the force components $\Delta f^{\mathrm{C1}_g}_{\mathrm{C1}_{x,y,z}}$ and of the force vector $||\Delta \mathbf{f}^{\mathrm{C1}_g}_{\mathrm{C1}}||$ of the reference atom $\mathrm{C1}$ (see Fig.~\ref{fig:FDHessianC1}) for different fragments in $\mathrm{eV\,\text{\AA{}}^{-1}}$.
Further, the effective Hessian group matrix norm $||\mathbf{G'}_{\mathrm{C1}}^{g}||$ is given in $\mathrm{eV\,\text{\AA{}}^{-2}}$. Numbers outside the intended convergence level are given in bold. The fragments C1$_\mathrm{1}$ - C1$_\mathrm{7}$ are shown in Fig.~S9 
.}
     \label{tab:SC_C1}
     \begin{ruledtabular}
     \begin{tabular}{ccccccc}
        $g$	&	$\Delta f^{\mathrm{C1}_g}_{\mathrm{C1}_{x}}$	&	$\Delta f^{\mathrm{C1}_g}_{\mathrm{C1}_{y}}$	&	$\Delta f^{\mathrm{C1}_g}_{\mathrm{C1}_{z}}$	    & $||\Delta\mathbf{f}^{\mathrm{C1}_g}_{\mathrm{C1}}||$	    &	$||\mathbf{G}'^{g}_{\mathrm{C1}}||$	\\
\hline
ref	&	0.0000	&	0.0000	&	0.0000	&	0.0000	&	0.00							\\
1	&	-0.0018	&	0.0022	&	0.0165	&	0.0167	&	0.02							\\
2	&	-0.0019	&	0.0022	&	-0.0733	&	0.0734	&	0.10							\\
3	&	-0.0019	&	0.0022	&	-0.0630	&	0.0630	&	0.11							\\
4	&	-0.0013	&	0.0016	&	-0.0599	&	0.0599	&	$\mathbf{0.36}$							\\
5	&	-0.0024	&	0.0027	&	0.0172	&	0.0176	&	0.35							\\
6	&	-0.0018	&	0.0022	&	$\mathbf{0.1666}$	&	$\mathbf{0.1666}$	&	0.30				\\
7	&	-0.0016	&	0.0016	&	$\mathbf{-3.0135}$	&	$\mathbf{3.0135}$	&	$\mathbf{21.83}$	\\
     \end{tabular}
     \end{ruledtabular}
\end{table}

The data for the other atomic sites in MOF-5 compiled in Tab.~S2
show a similar behavior and also here the application of the threshold yields fragments with well-converged forces. Still, the size of the resulting fragments as shown in Fig.~S16
can be very different depending on  the reference site reflecting the different bonding and local electronic structure. This has the interesting consequence that the converged fragment for reference site C1, i.e. C1$_3$ (see Figs.~ S9
 and S16) and O1 (Fig.~S16)
 effectively contain all the other fragments and thus at the same time can provide converged DFT forces for all the atomic sites. Thus, in principle a data set could be constructed using different configurations based on these fragments only, obtained e.g. by molecular dynamics simulations. For the construction of MLPs, however, this is not a viable approach, since the effective environment radius covered by this fragment is different for each reference site, while MLPs require the use of a consistent environment radius for all atoms of all elements in the systems. This radius has to be chosen as the largest environment radius in the converged fragments of all sites.

\begin{figure*}[!ht]
    \centering
    \includegraphics[width=\linewidth]{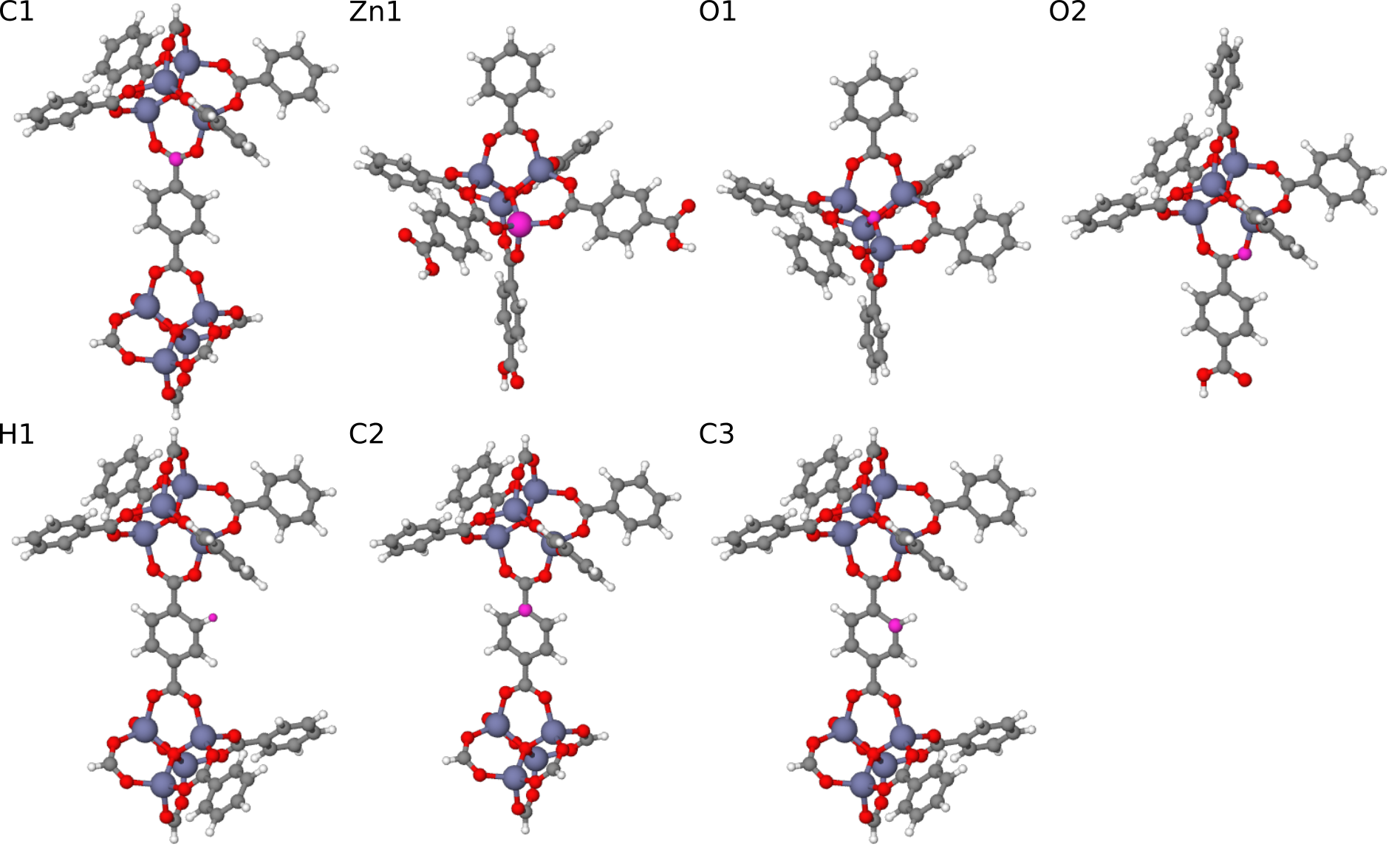}
    \caption{Fragments for the atomic sites C1, Zn1, O1, O2, H1, C2 and C3 (magenta) obtained with the common fragment radius $r_\mathrm{frag} = 8.502\,\text{\AA{}}$ (Tab.~\ref{tab:smallest-frags-distances}). The fragments of Zn1 and H1 or C3 effectively include all other fragments.}
    \label{fig:frags-FragmentRadius}
\end{figure*}

There are several possibilities to define such a radius for a given fragment. An obvious choice would be the distance of the outermost atom from the central reference atom, but in many cases we find that these atoms are only weakly interacting and have only been included, because they are part of structural entities like phenyl groups that cannot be cut in the fragment construction (see e.g. Fig. \ref{fig:FDHessianC1}a). Therefore, in the present work we define the fragment radius $r_\mathrm{frag}$ as the distance between the reference atom and the atom with the largest atomic Hessian submatrix norm value in the most distant functional group of the fragment. The resulting fragment radii are compiled in Table \ref{tab:smallest-frags-distances}. The largest radius is $r_\mathrm{frag}=8.502$~\AA{} for the environment of C1, which we accordingly select as common fragment radius for all atomic environments resulting in a new set of fragments shown in Fig. \ref{fig:frags-FragmentRadius}. 

\begin{figure*}
    \centering
    \includegraphics{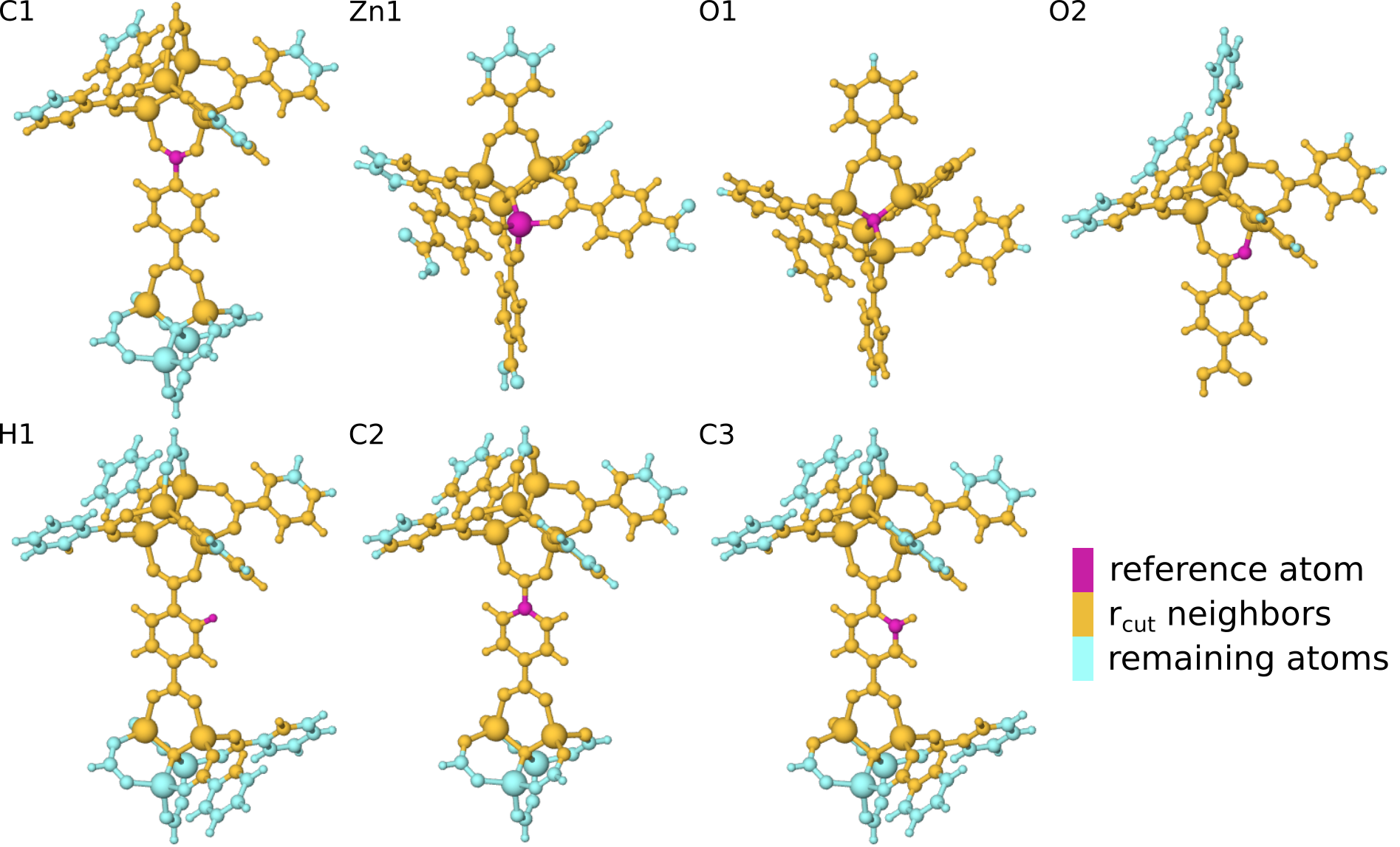}
    \caption{Atomic neighbors of the reference atoms C1, Zn1, O1, O2, H1, C2 and C3 (magenta) within a radius $r_\mathrm{frag} = 8.502\,$\AA{} (orange) and atoms needed in addition to complete the outermost functional groups that cannot be cut (blue).}
    \label{fig:851neighbors}
\end{figure*}

An inspection of these structures shows that the fragments for Zn1 and for H1 or C3, which are identical, include the fragments of C1, O1, O2 and C2 such that only two unique fragments are required to construct the training set for MLPs, which offers the advantage of obtaining converged reference data for different atomic sites in the same DFT calculations. Moreover we note that the resulting fragments are not sensitive to the specific definition of the fragment radii in Table~\ref{tab:smallest-frags-distances}, because in any case the main structural building blocks like SBUs, carboxyl and phenyl groups cannot be cut and have to be included completely. Figure \ref{fig:851neighbors} shows all atoms within the common fragment radius (orange), which are the atoms that would determine the atomic energy of the reference atoms (magenta) in MLPs employing the fragment radius as cutoff radius, while the atoms colored in blue are only required to complete the outermost functional groups to avoid significant changes of the electronic structure within the fragment radius.

Finally we note that in principle a more efficient construction of the training sets might be possible when increasing the number of atoms in a bulk-like environment by slightly increasing the fragment radius. For instance, in the Zn1 fragment in Fig. \ref{fig:frags-FragmentRadius} only one of the four Zn atoms (magenta) has a bulk-like environment using the determined fragment radius $r_\mathrm{frag}=8.502\,$\AA{}, while in the environment of the other three zinc atoms of the SBU the carboxyl goups terminating the phenyl groups are missing. Adding these groups would thus strongly increase the amount of information of the DFT calculation at only a moderate increase in the number of atoms. 

\begin{table}[!ht]
    \centering
    \caption{Fragment radii $r_\mathrm{frag}$ in \AA{} obtained for the fragments shown in Fig.~S16
    for the seven atomic sites in MOF-5. $g$ is the number of the converged fragment.}
    \label{tab:smallest-frags-distances}
    \begin{ruledtabular}
    \begin{tabular}{ccc}
        $A$ &   $g$ &   $r_\mathrm{frag}$ \\
\hline
        Zn1 &   3   &   4.333   \\
        O1  &   3   &   5.165   \\
        O2  &   6   &   2.379   \\
        C1  &   3   &   8.502   \\
        C2  &   3   &   7.304   \\
        C3  &   4   &   3.817   \\
        H1  &   5   &   2.725   \\
    \end{tabular}
    \end{ruledtabular}
\end{table}

\section{Conclusions}\label{sec:conclusions}

In this work we have presented a locality test based on the Hessian to determine size-converged molecular fragments providing reliable atomic forces suitable for training machine learning interatomic potentials. A set of one-dimensional model systems representing different types of bonding has been employed to identify a suitable threshold value to assess the joint effect of  groups of atoms as a function of distance. We have found that irrespective of the chemical bonding situation very similar thresholds for the effective Hessian group matrix norm are obtained, which are to a good approximation transferable across the investigated model systems. Moreover, the applicability of these thresholds to the more complex metal-organic-framework MOF-5 has been explored in one and three dimensions confirming that molecular fragments with well converged forces can be obtained. A recipe has been given to construct molecular fragments centered at different atomic sites with uniform fragment radius suitable for the construction of machine learning potentials. The method is general and also applicable to rather symmetric, e.g. crystalline, environments, which are difficult to investigate by monitoring the force convergence directly. Moreover, our approach allows to determine the degree of locality of the atomic interactions and to identify situations in which interactions beyond the local atomic environments may have to be included using, e.g., third or fourth-generation MLPs, due to the importance of long-range interactions like electrostatics.

\begin{acknowledgments}
We thank the Deutsche Forschungsgemeinschaft (DFG) for financial support (BE3264/12-1, project number 405479457 as part of PAK 965/1). We gratefully acknowledge computing time provided by the DFG project INST186/1294-1 FUGG (Project No.\ 405832858). 
\end{acknowledgments}

 \bibliography{literature}

\end{document}


\title{A Hessian-Based Assessment of Atomic Forces for Training Machine Learning Interatomic Potentials}

\author{Marius Herbold}
\author{J\"{o}rg Behler}
\email{joerg.behler@uni-goettingen.de}
\affiliation{Universit\"{a}t G\"{o}ttingen, Institut f\"{u}r Physikalische Chemie, Theoretische Chemie, Tammannstra\ss{}e 6, 37077 G\"{o}ttingen, Germany}

\maketitle

\onecolumngrid
\setcounter{table}{0}
\setcounter{figure}{0}
\setcounter{equation}{0}
\setcounter{section}{0}
\renewcommand{\thetable}{S\arabic{table}}%
\renewcommand{\thefigure}{S\arabic{figure}}%
\renewcommand{\theequation}{S\arabic{equation}}
\renewcommand{\thesection}{S-\Roman{section}}

\section{Model Systems}
\begin{figure*}[!htbp]
    \centering
    \includegraphics[scale=0.95]{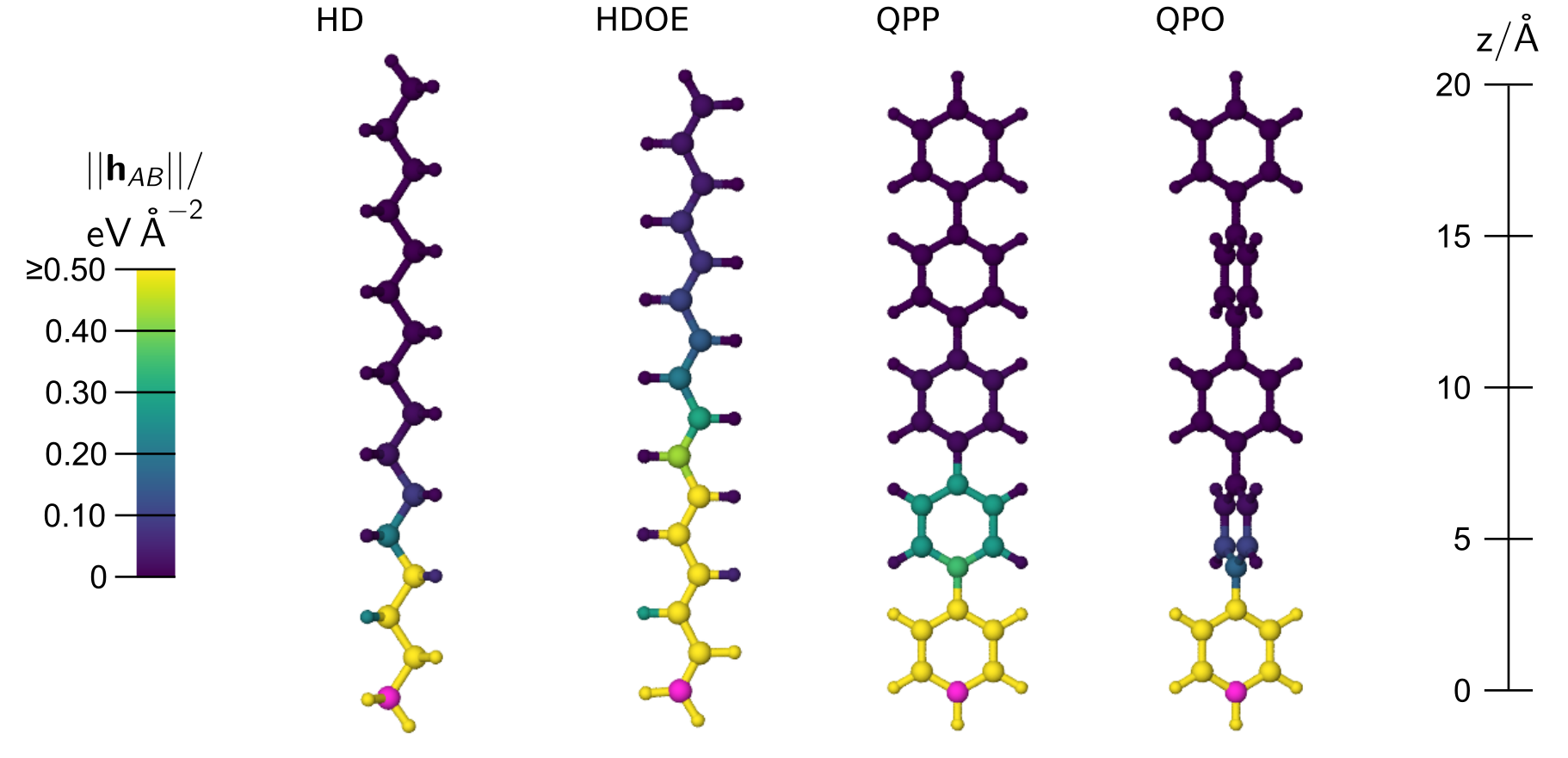}
    \caption{Atomic Hessian submatrix norm values $||\mathbf{h}_{AB}||$ describing the interaction between the magenta reference carbon atoms $A$ and all other atoms $B$ in the model systems HD, HDOE, QPP and QPO (linear scale).}
    \label{fig:Model_System_HessianSubMatrixNorm}
\end{figure*}

\begin{figure*}[!ht]
    \centering
    \includegraphics{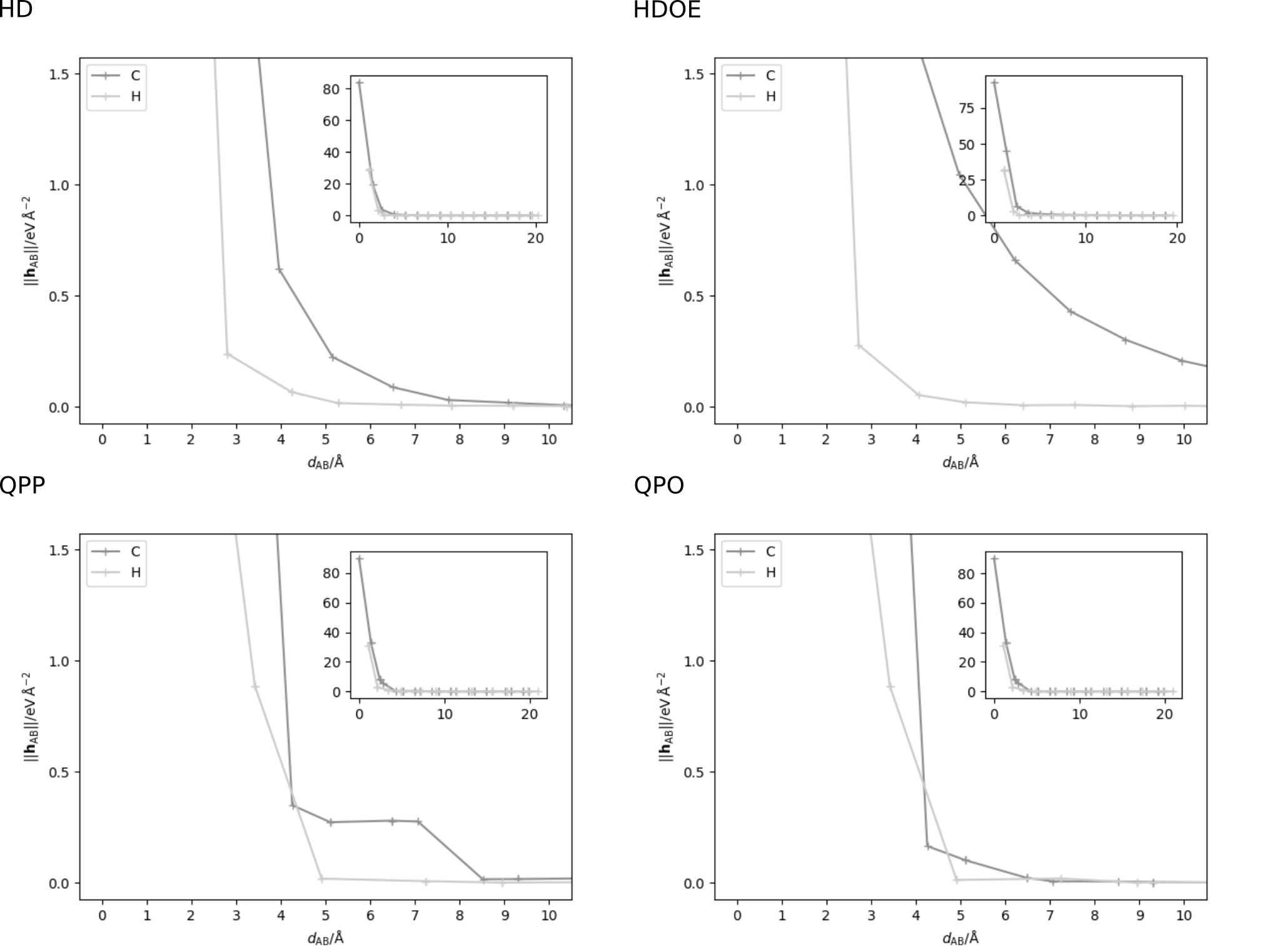}
    \caption{Atomic Hessian submatrix norm values $||\mathbf{h}_{AB}||$ of the four model systems HD, HDOE, QPP and QPO as a function of the distance $d_{AB}$ between the reference atom $A$ as defined in Fig.~4
    and atoms $B$. Separate curves are given for the interactions of atom $A$ with neighboring carbon and hydrogen atoms. The inset shows the data for the interaction of $A$ with all atoms in the entire molecules.}
    \label{fig:Model_System_distanceVSnorm_separated}
\end{figure*}

\begin{figure*}[!ht]
    \centering
    \includegraphics{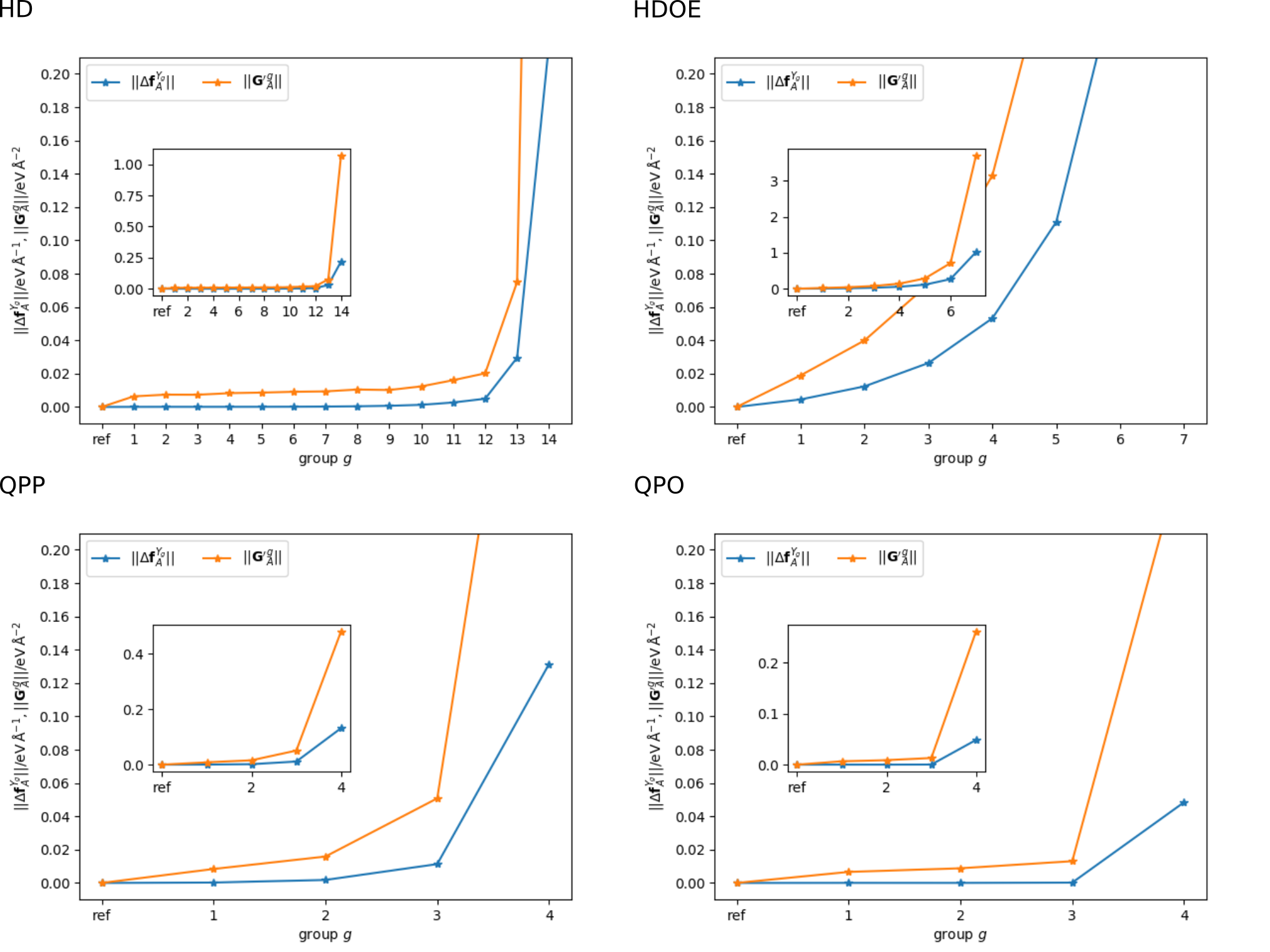}
    \caption{Effective Hessian group matrix norm values $||\mathbf{G'}_{A}^g||$ of group $g$ and the norm of the force error $||\Delta \mathbf{f}^{Y_g}_A||$ of the reference carbon atom $A$ in the fragment structure $Y_g$ (reference structure $Y$ without the removed atoms of group $g$, including hydrogen saturation) for the model systems HD, HDOE, QPP and QPO. The force difference vector is defined as $\Delta \mathbf{f}^{Y_g}_A = \mathbf{f}^{Y}_A - \mathbf{f}^{Y_g}_A$.}
    \label{fig:Model_System_F+HoverGROUP}
\end{figure*}

\begin{figure}
    \centering
    \includegraphics[scale=0.6]{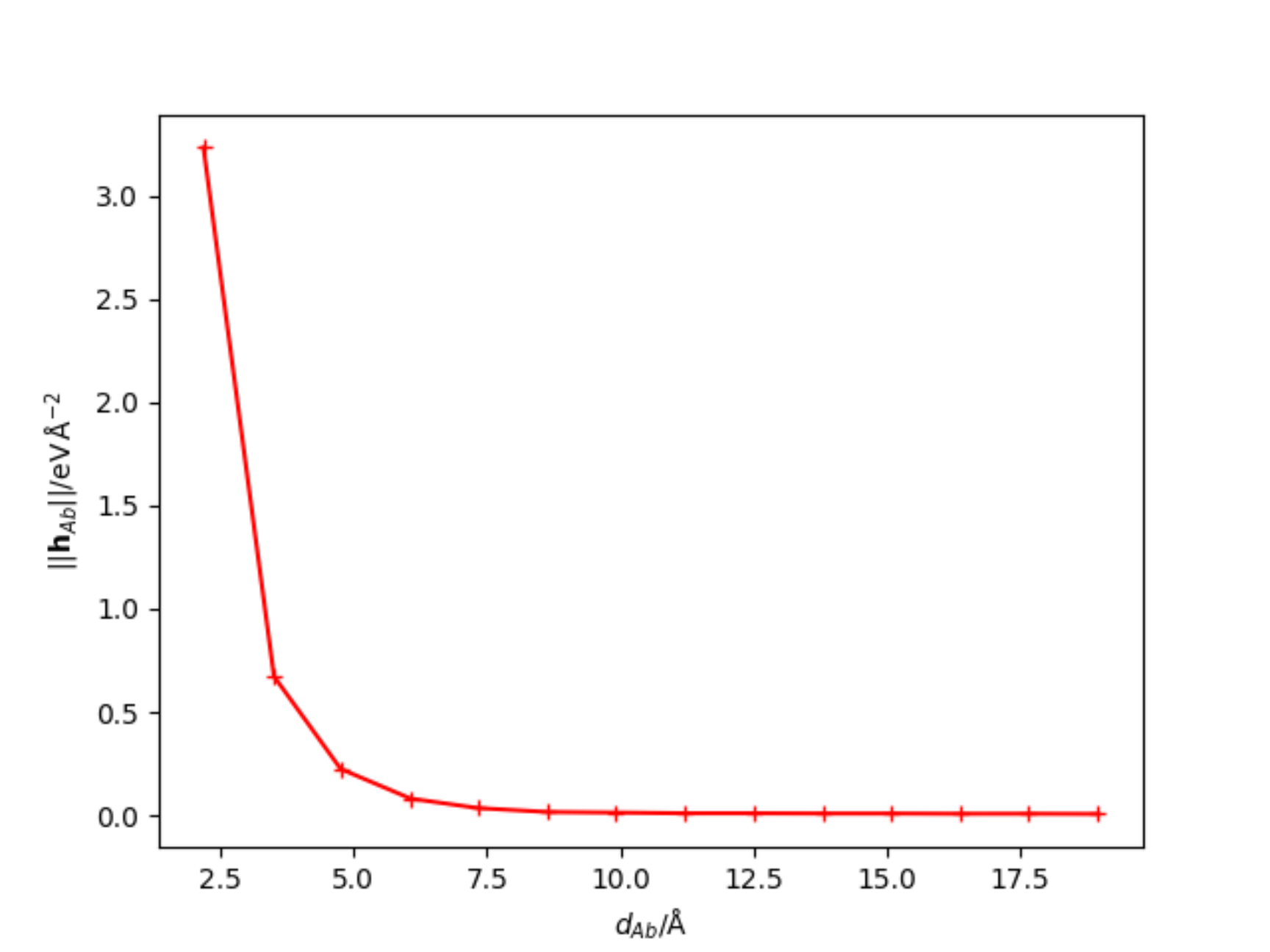}
    \caption{Atomic Hessian sub matrix norm values $||\mathbf{h}_{Ab}||$ for the saturating hydrogen atom $b$ as a function of the distance $d_{Ab}$ to the reference carbon atom $A$ in the different fragments of the HD model system.}
    \label{fig:Model_System_SaturationHydrogen}
\end{figure}

\begin{table*}[!ht]
     \centering
     \caption{
     Compilation of the force component errors $\Delta f^{Y_g}_{A_{x,y,z}}$ and the total force errors $||\Delta \mathbf{f}^{Y_g}_{A}||$ in $\mathrm{eV\,\text{\AA{}}^{-1}}$ for the reference carbon atoms in the model systems $Y=\mathrm{HD}, \mathrm{HDOE}, \mathrm{QPP}, \mathrm{QPO}$ (Fig.~6
     and 7,
     $\sigma=1.00$). Further, the effective Hessian group matrix norm $||\mathbf{G'}_{A}^{g}||$ is given in $\mathrm{eV\,\text{\AA{}}^{-2}}$. Numbers outside the intended convergence are given in bold.}
     \label{tab:SC_Model_Systems}
     \begin{ruledtabular}
     \begin{tabular}{ccccccc}
        $Y$ & $g$	&	$\Delta f^{Y_g}_{A_\mathrm{x}}$	&	$\Delta f^{Y_g}_{A_\mathrm{y}}$	&	$\Delta f^{Y_g}_{A_\mathrm{z}}$ & $||\Delta\mathbf{f}^{Y_g}_{A}||$	    &	$||\mathbf{G}'^{g}_{A}||$	\\ \hline
HD	&	ref	&	0.0000	&	0.0000	&	0.0000	&	0.0000	&	0.00    \\
	&	1	&	0.0000	&	0.0000	&	0.0000	&	0.0000	&	0.01    \\
	&	2	&	0.0000	&	$-0.0001$	&	0.0000	&	0.0001	&	0.01    \\
	&	3	&	0.0000	&	$-0.0001$	&	0.0000	&	0.0001	&	0.01    \\
	&	4	&	0.0000	&	$-0.0001$	&	0.0000	&	0.0001	&	0.01    \\
	&	5	&	0.0000	&	$-0.0001$	&	0.0000	&	0.0001	&	0.01    \\
	&	6	&	0.0000	&	$-0.0001$	&	0.0000	&	0.0001	&	0.01    \\
	&	7	&	0.0000	&	$-0.0002$	&	0.0000	&	0.0002	&	0.01    \\
	&	8	&	0.0000	&	$-0.0004$	&	$-0.0001$	&	0.0004	&	0.01    \\
	&	9	&	0.0000	&	$-0.0007$	&	$-0.0003$	&	0.0007	&	0.01    \\
	&	10	&	0.0000	&	$-0.0011$	&	$-0.0006$	&	0.0013	&	0.01    \\
	&	11	&	0.0000	&	$-0.0025$	&	$-0.0009$	&	0.0027	&	0.02    \\
	&	12	&	0.0000	&	$-0.0039$	&	$-0.0030$	&	0.0049	&	0.02    \\
	&	13	&	0.0000	&	0.0054	&	$-0.0291$	&	0.0296	&	0.08    \\
	&	14	&	0.0000	&	$\mathbf{-0.1981}$	&	-0.0921	&	$\mathbf{0.2184}$	&	$\mathbf{1.07}$			\\
HDOE	&	ref	&	0.0000	&	0.0000	&	0.0000	&	0.0000	&	0.00	\\
	&	1	&	0.0000	&	0.0014	&	$-0.0043$	&	0.0045	&	0.02	\\
	&	2	&	0.0000	&	0.0042	&	$-0.0116$	&	0.0124	&	0.04	\\
	&	3	&	0.0000	&	0.0094	&	$-0.0247$	&	0.0264	&	0.07	\\
	&	4	&	0.0000	&	0.0195	&	$-0.0495$	&	0.0532	&	0.14	\\
	&	5	&	0.0000	&	0.0414	&	$-0.1031$	&	0.1111	&	0.29	\\
	&	6	&	0.0000	&	0.0986	&	$\mathbf{-0.2474}$	&	$\mathbf{0.2663}$	&	$\mathbf{0.71}$		\\
	&	7	&	0.0000	&	$\mathbf{0.1877}$	&	$\mathbf{-1.0018}$	&	$\mathbf{1.0192}$	&	$\mathbf{3.71}$		\\
QPP	&	ref	&	0.0000	&	0.0000	&	0.0000	&	0.0000	&	0.00		\\
	&	1	&	0.0000	&	0.0000	&	$-0.0003$	&	0.0003	&	0.01		\\
	&	2	&	0.0000	&	0.0000	&	$-0.0018$	&	0.0018	&	0.02		\\
	&	3	&	0.0000	&	0.0000	&	$-0.0113$	&	0.0113	&	0.05		\\
	&	4	&	0.0000	&	0.0000	&	$-0.1315$	&	0.1315	&	$\mathbf{0.48}$			\\
QPO	&	ref	&	0.0000	&	0.0000	&	0.0000	&	0.0000	&	0.00			\\
	&	1	&	0.0000	&	0.0000	&	0.0000	&	0.0000	&	0.01			\\
	&	2	&	0.0000	&	0.0000	&	0.0000	&	0.0000	&	0.01			\\
	&	3	&	0.0000	&	0.0000	&	$-0.0002$	&	0.0002	&	0.01    		\\
	&	4	&	0.0000	&	0.0000	&	$-0.0484$	&	0.0484	&	0.26			\\
     \end{tabular}
     \end{ruledtabular}
 \end{table*}

\clearpage
\section{MOF-5 Fragment Construction}
\label{sec:frag_const}
 
For the construction of the MOF-5 fragments the electronic structure should be kept as close as possible to bulk MOF-5. To achieve this goal, we define several functional groups of atoms, which cannot be cut in the fragment construction, which are the SBUs, phenylene rings and carboxyl groups. To avoid terminating zinc and oxygen atoms in the $\mathrm{Zn_4O}$ tetrahedra of the SBUs, we consider an SBU as an entity including six formate species.
Consequently, the MOF-5 fragments can be terminated by phenyl groups, carboxyl groups or formate-capped SBUs, and there are three possibilities to break bonds: 1. between the carboxyl group and the phenylene ring, 2. between the phenylene ring and the carboxyl group and 3. between the carboxyl group and the attached neighboring SBU.

\clearpage
\section{MOF-5 1D-Structures}

 \begin{figure*}[!ht]
    \centering
    \includegraphics{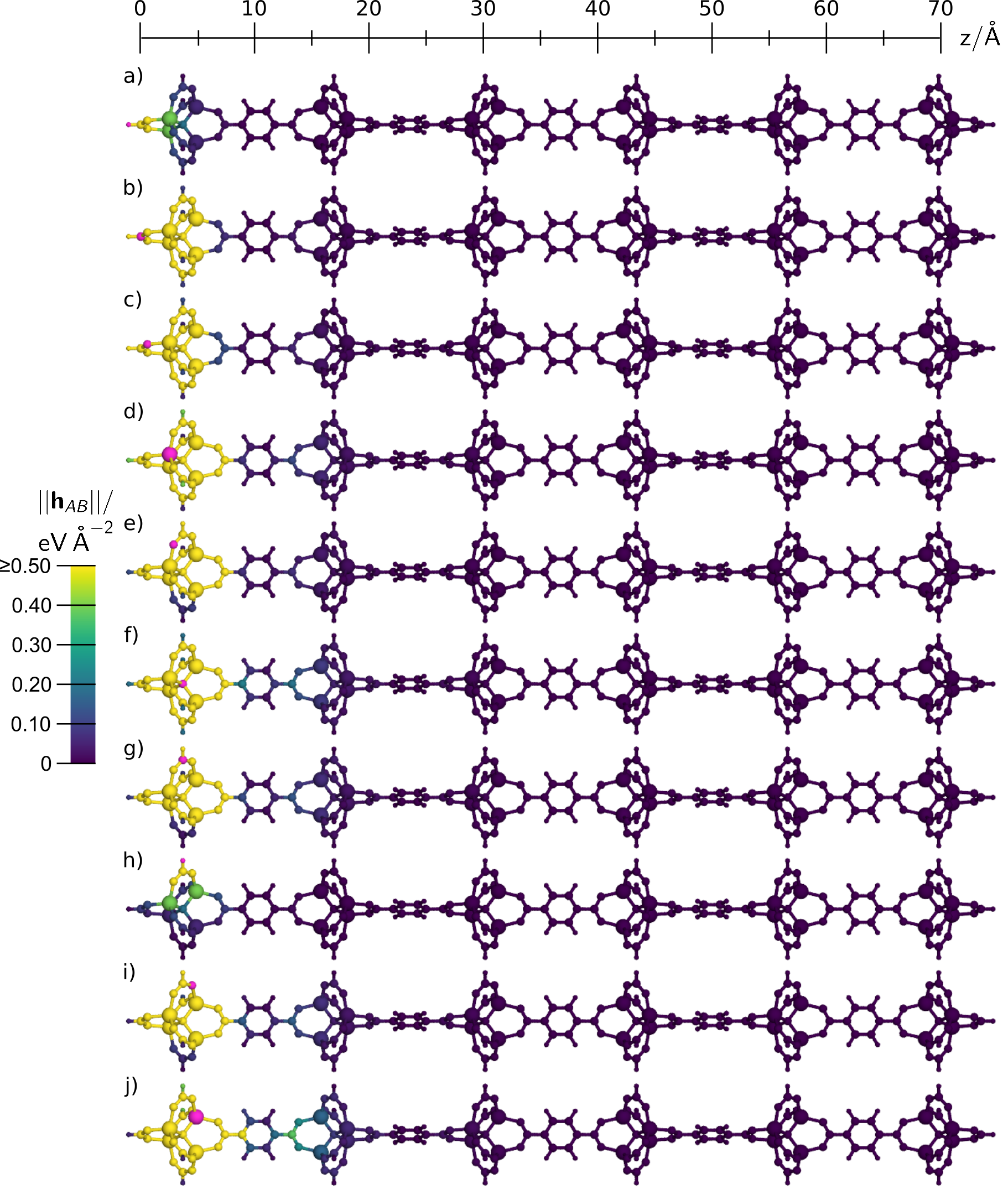}
    \caption{a)--j) Atomic Hessian submatrix norm values $||\mathbf{h}_{AB}||$ for different reference atoms $A$ (magenta) of the one-dimensional reference structure 1D of MOF-5.}
\label{fig:1D-MOF_Hessian_atomic_norms}
\end{figure*}

\begin{figure*}[!ht]
    \centering
    \includegraphics{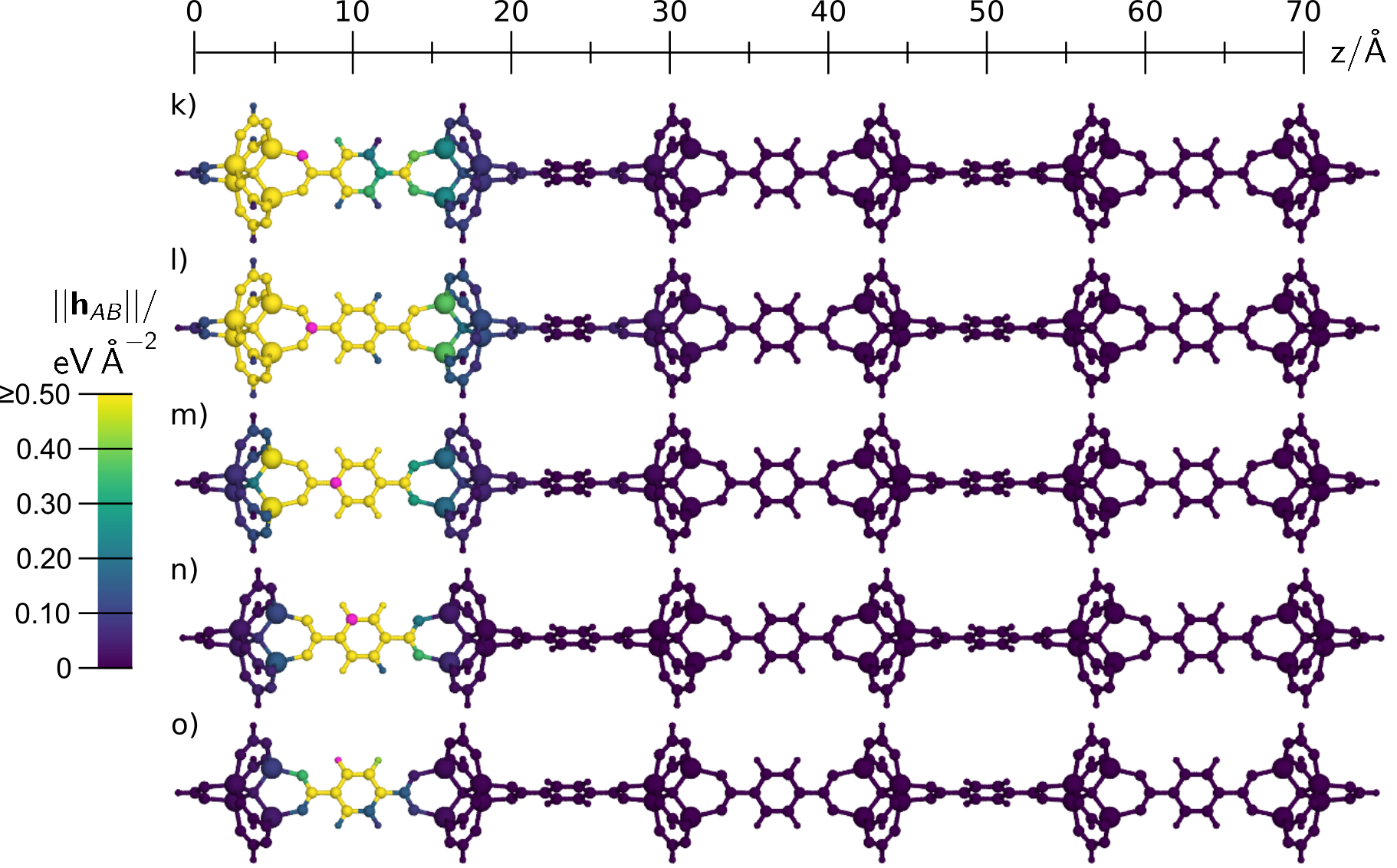}
    \caption{Continuation of Fig. \ref{fig:1D-MOF_Hessian_atomic_norms}. k)--o) Atomic Hessian submatrix norm values $||\mathbf{h}_{AB}||$ for different reference atoms $A$ (magenta) of the one-dimensional reference structure 1D of MOF-5.}
\label{fig:1D-MOF_Hessian_atomic_norms_continuation}
\end{figure*}

\begin{figure*}[!ht]
    \centering
    \includegraphics{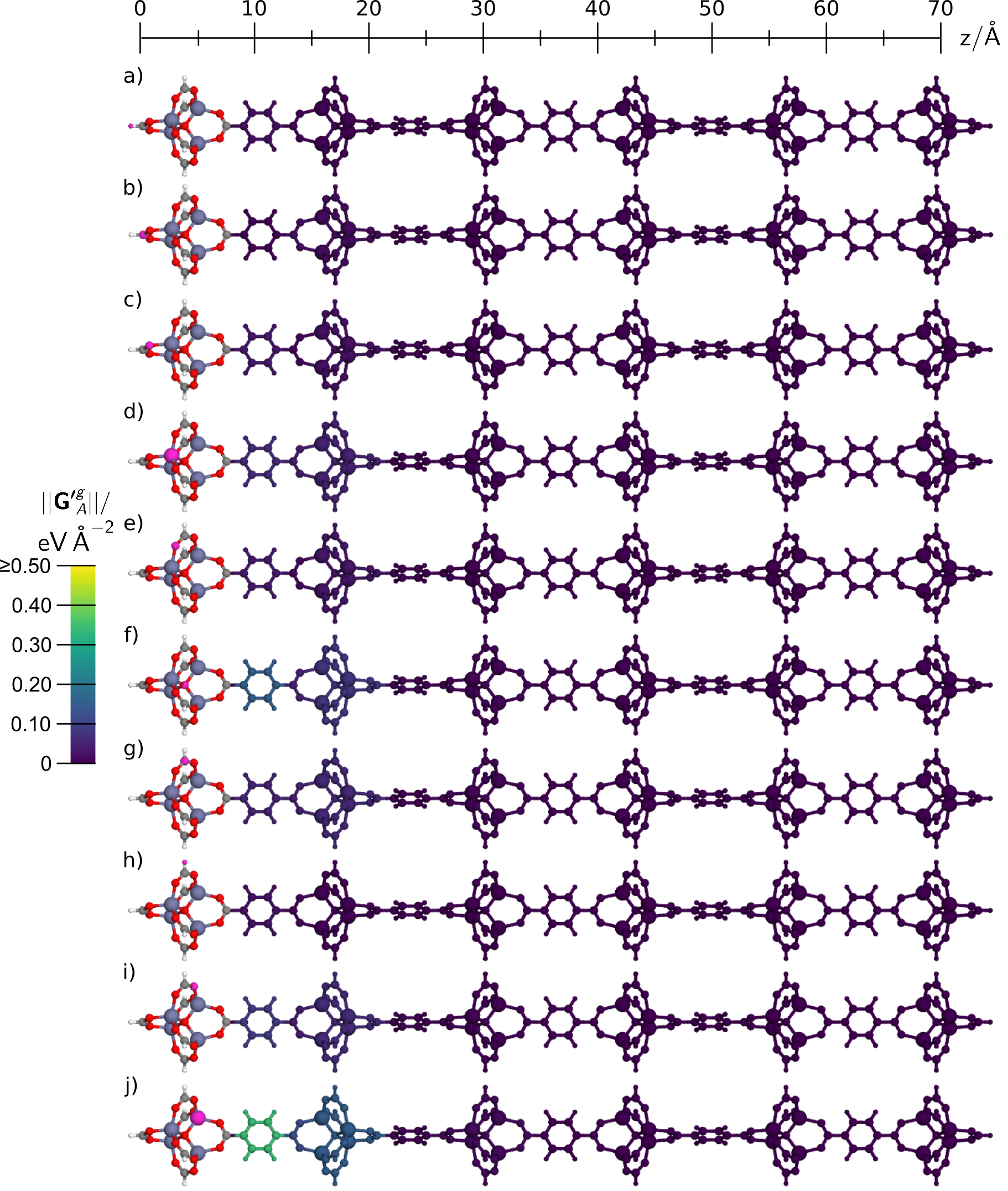}
    \caption{a)--j) Effective Hessian group matrix norm values $||\mathbf{G'}^{g}_A||$ for different reference atoms $A$ (magenta) of the one-dimensional reference structure 1D of MOF-5. The atomic colors of the smallest possible fragment are specified by the atom's element.}
\label{fig:1D-MOF_Hessian_extGROUP_norms}
\end{figure*}

\begin{figure*}[!ht]
    \centering
    \includegraphics{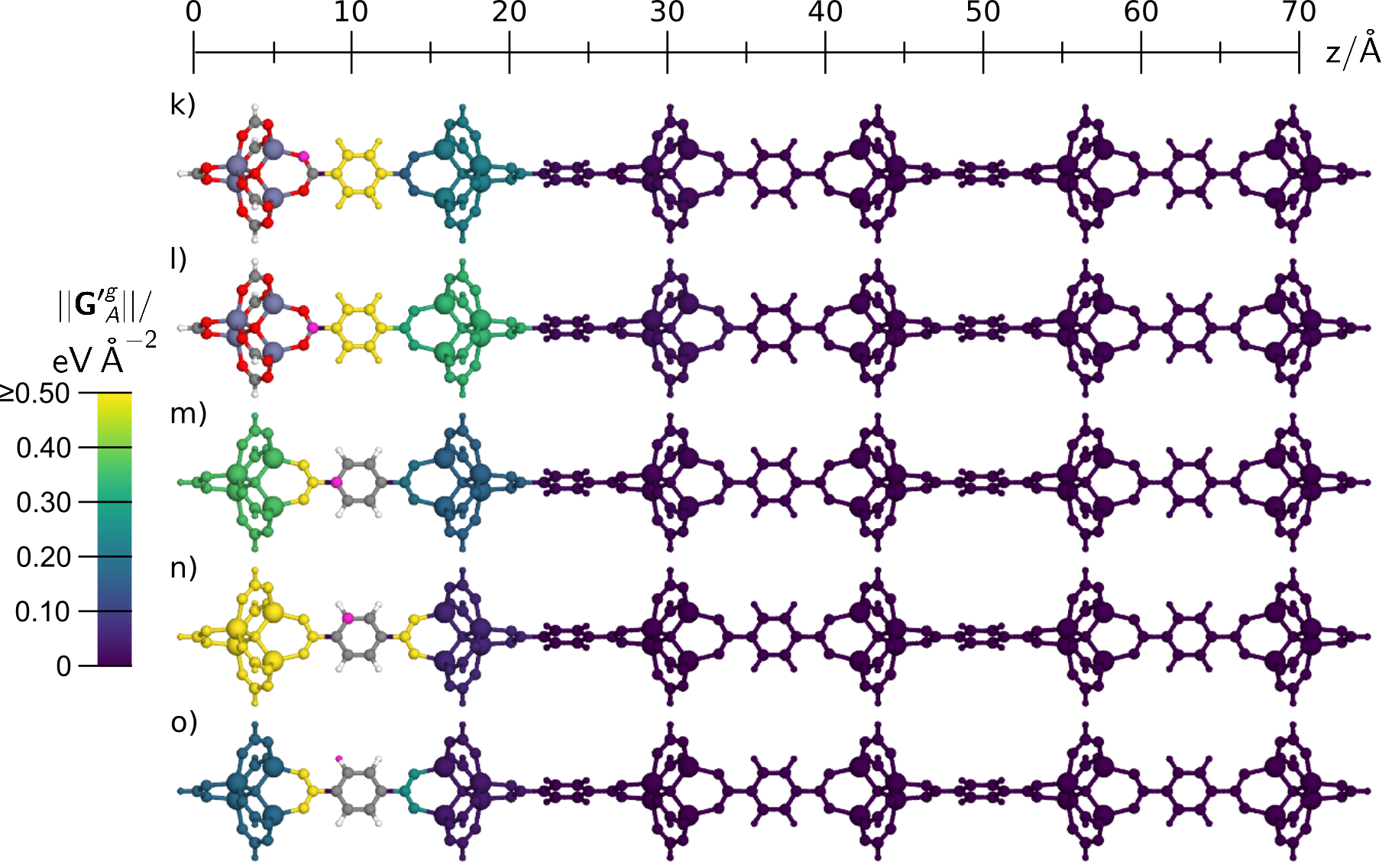}
    \caption{Continuation of Fig. \ref{fig:1D-MOF_Hessian_extGROUP_norms}. k)--o) Effective Hessian group matrix norm values $||\mathbf{G'}^{g}_A||$ for different reference atoms $A$ (magenta) of the one-dimensional reference structure 1D of MOF-5. The atomic colors of the smallest possible fragment are specified by the atom's element.}
\label{fig:1D-MOF_Hessian_extGROUP_norms_continuation}
\end{figure*}

\clearpage
\newpage
\section{IRMOF 3D-Fragment Structures}

\begin{figure*}[!ht]
    \centering
    \includegraphics{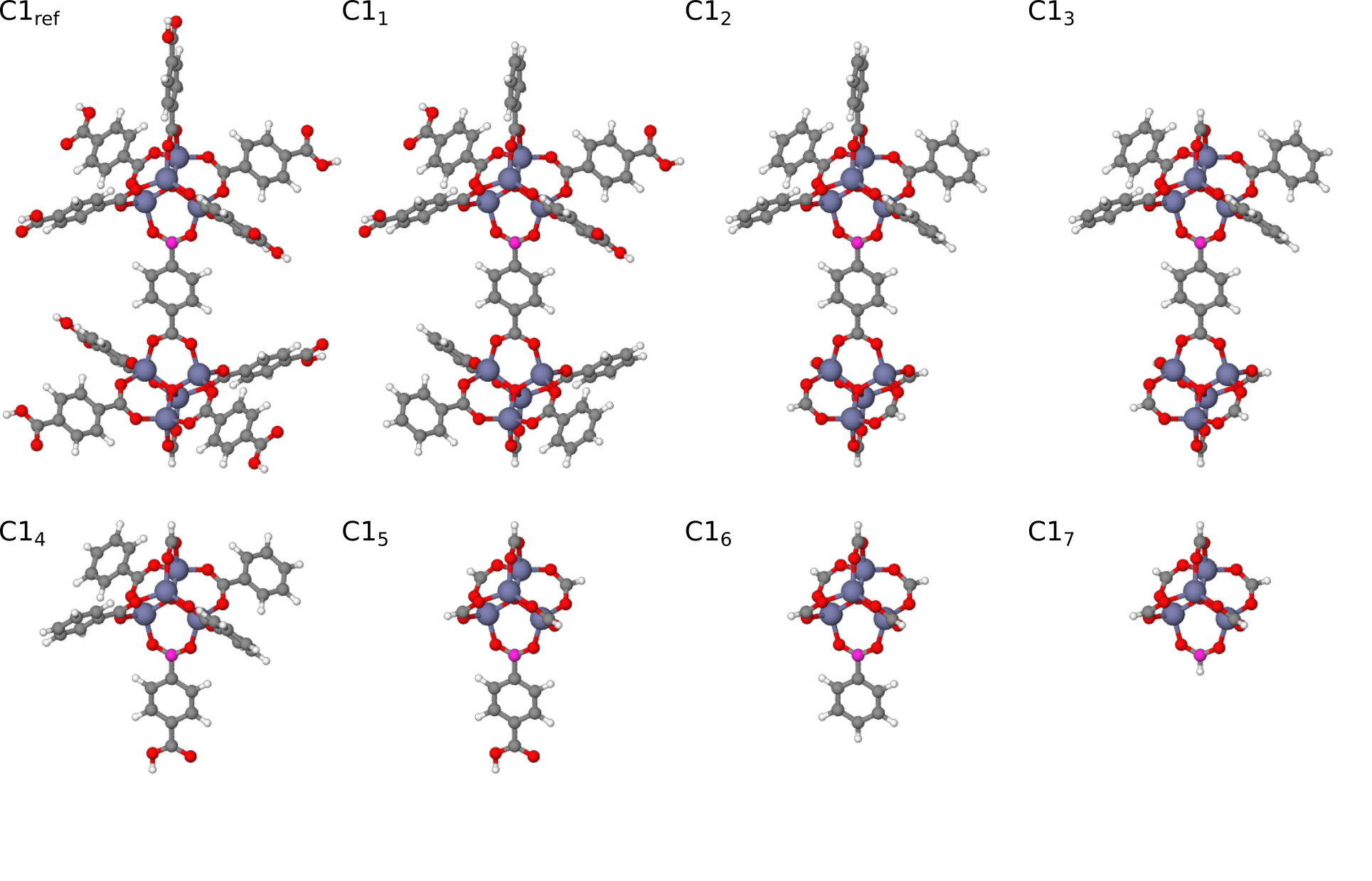}
    \caption{The reference structure C1$_\mathrm{ref}$ and the molecular fragments C1$_\mathrm{1}$ to C1$_\mathrm{7}$ constructed for the reference site C1 (magenta) in MOF-5 including the hydrogen atoms saturating the broken bonds.}
    \label{fig:C1-Fragments}
\end{figure*}

\begin{figure*}[!hp]
    \centering
    \includegraphics[scale=1.0]{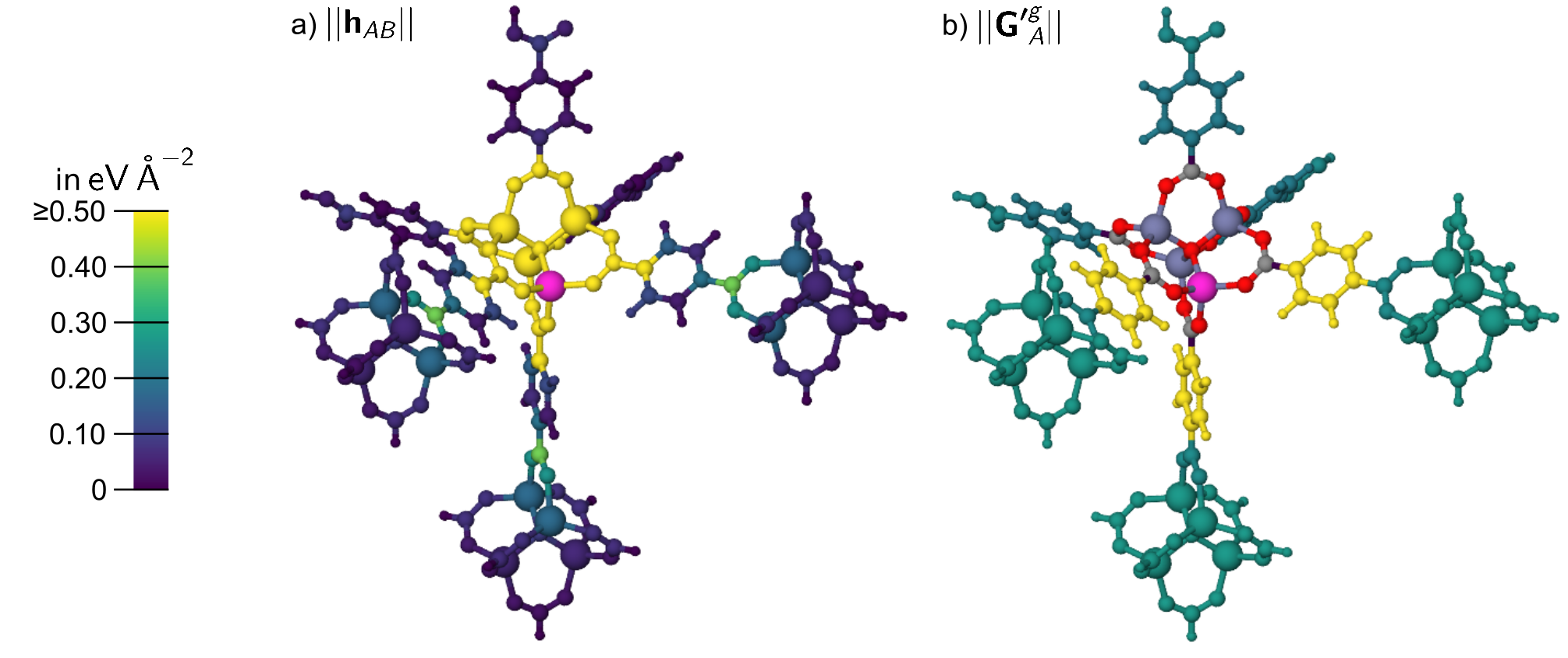}
    \caption{a) Atomic Hessian submatrix norm values $||\mathbf{h}_{AB}||$ and b) effective Hessian group matrix norm values $||\mathbf{G'}_{A}^g||$ in $\text{eV\,\AA{}}^{-2}$ with respect to the central atom $A=\mathrm{Zn1}$ (magenta) in reference structure $\mathrm{Zn1_{ref}}$. $||\mathbf{G'}_{A}^g||$ defines the color for the closest atoms of a given group, which in addition also contains all atoms at larger distance. The colors of the smallest possible fragment in b) refer to the chemical elements.}
    \label{fig:FDHessianZn1}
\end{figure*}

\begin{figure*}[!hp]
    \centering
    \includegraphics[scale=1.0]{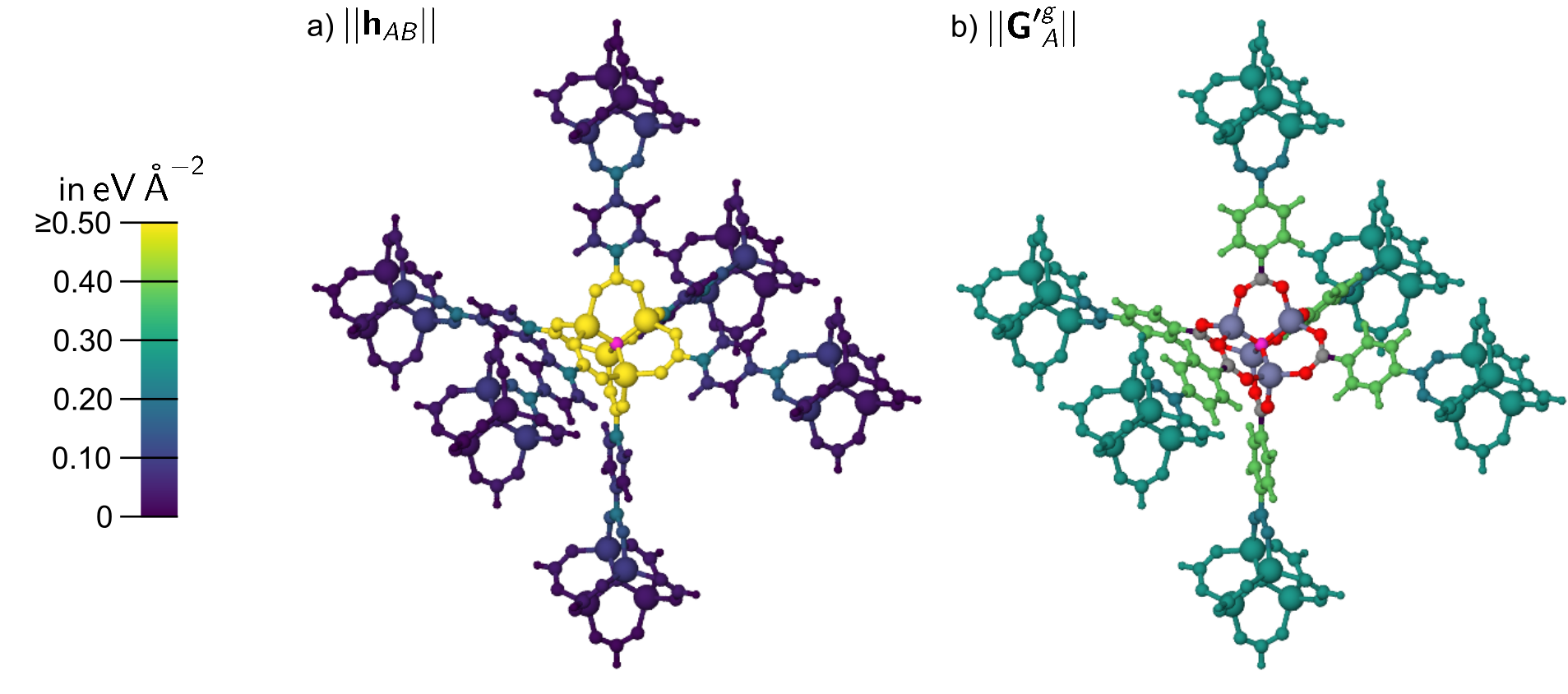}
    \caption{a) Atomic Hessian submatrix norm values $||\mathbf{h}_{AB}||$ and b) effective Hessian group matrix norm values $||\mathbf{G'}_{A}^g||$ in $\text{eV\,\AA{}}^{-2}$ with respect to the central atom $A=\mathrm{O1}$ (magenta) in reference structure $\mathrm{O1_{ref}}$. $||\mathbf{G'}_{A}^g||$ defines the color for the closest atoms of a given group, which in addition also contains all atoms at larger distance. The colors of the smallest possible fragment in b) refer to the chemical elements.}
    \label{fig:FDHessianO1}
\end{figure*}

\begin{figure*}[!hp]
    \centering
    \includegraphics[scale=1.0]{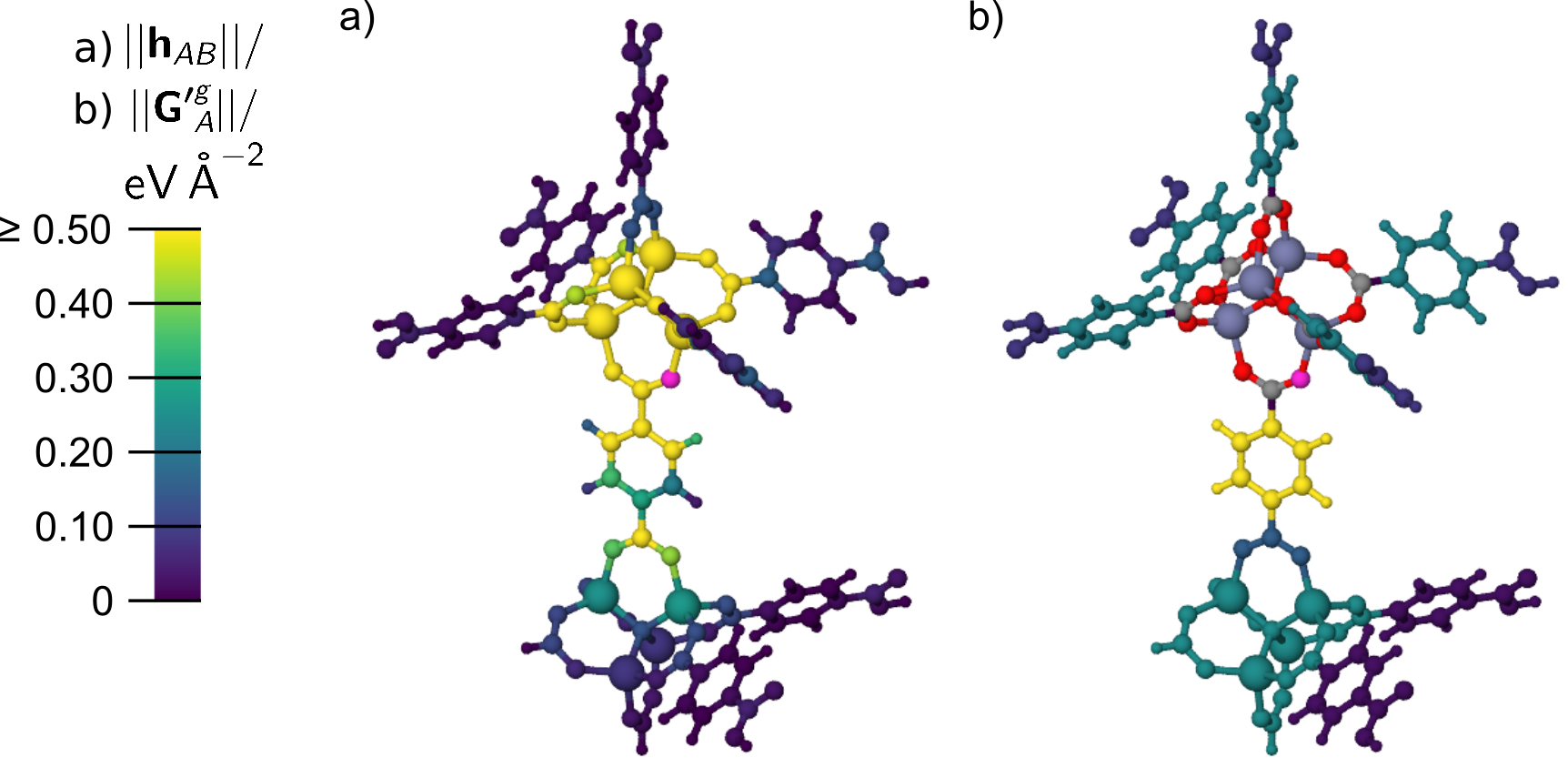}
    \caption{a) Atomic Hessian submatrix norm values $||\mathbf{h}_{AB}||$ and b) effective Hessian group matrix norm values $||\mathbf{G'}_{A}^g||$ in $\text{eV\,\AA{}}^{-2}$ with respect to the central atom $A=\mathrm{O2}$ (magenta) in reference structure $\mathrm{O2_{ref}}$. $||\mathbf{G'}_{A}^g||$ defines the color for the closest atoms of a given group, which in addition also contains all atoms at larger distance. The colors of the smallest possible fragment in b) refer to the chemical elements.}
    \label{fig:FDHessianO2}
\end{figure*}

\begin{figure*}[!hp]
    \centering
    \includegraphics[scale=1.0]{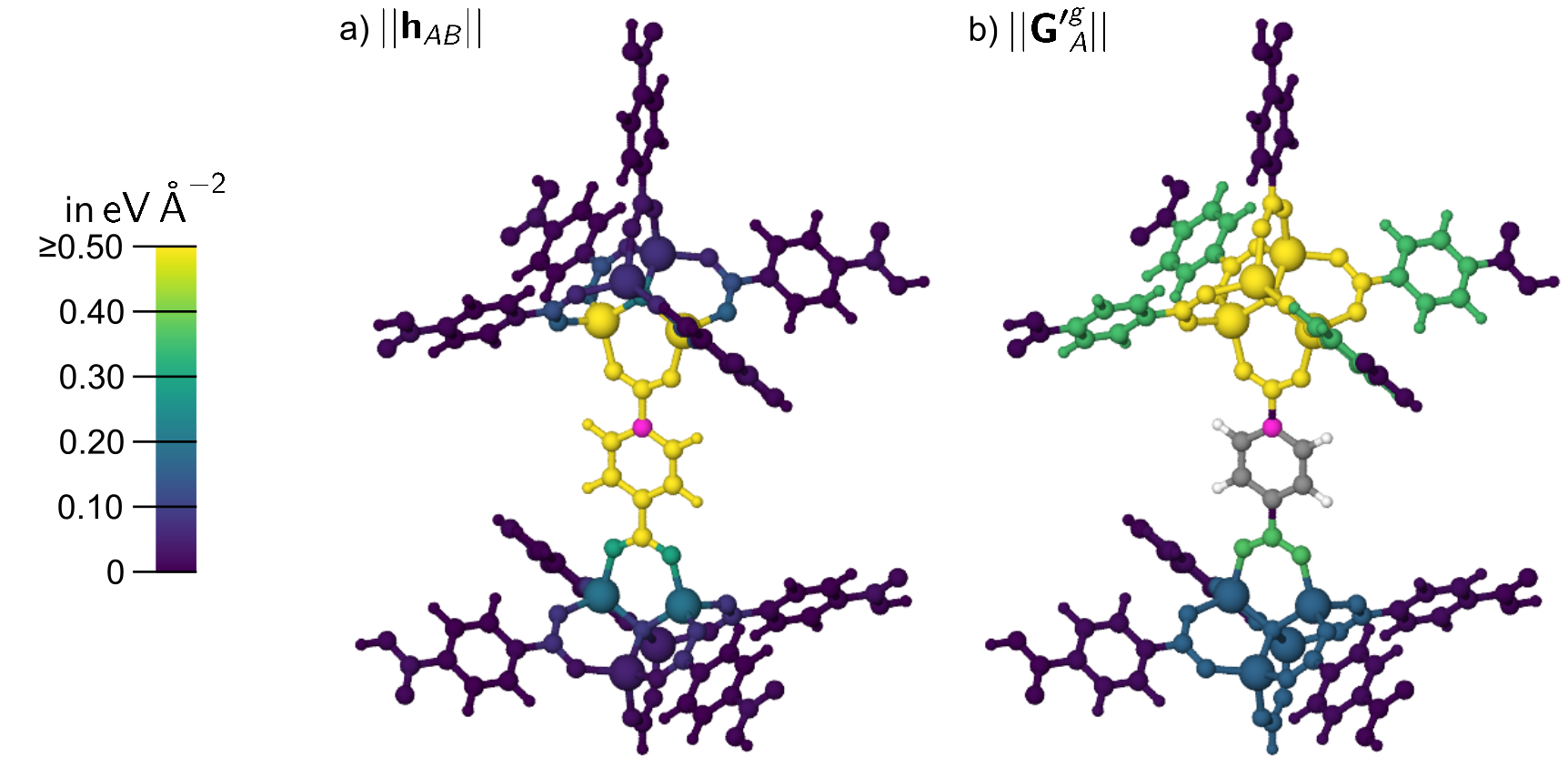}
    \caption{a) Atomic Hessian submatrix norm values $||\mathbf{h}_{AB}||$ and b) effective Hessian group matrix norm values $||\mathbf{G'}_{A}^g||$ in $\text{eV\,\AA{}}^{-2}$ with respect to the central atom $A=\mathrm{C2}$ (magenta) in reference structure $\mathrm{C2_{ref}}$. $||\mathbf{G'}_{A}^g||$ defines the color for the closest atoms of a given group, which in addition also contains all atoms at larger distance. The colors of the smallest possible fragment in b) refer to the chemical elements.}
    \label{fig:FDHessianC2}
\end{figure*}

\begin{figure*}[!hp]
    \centering
    \includegraphics[scale=1.0]{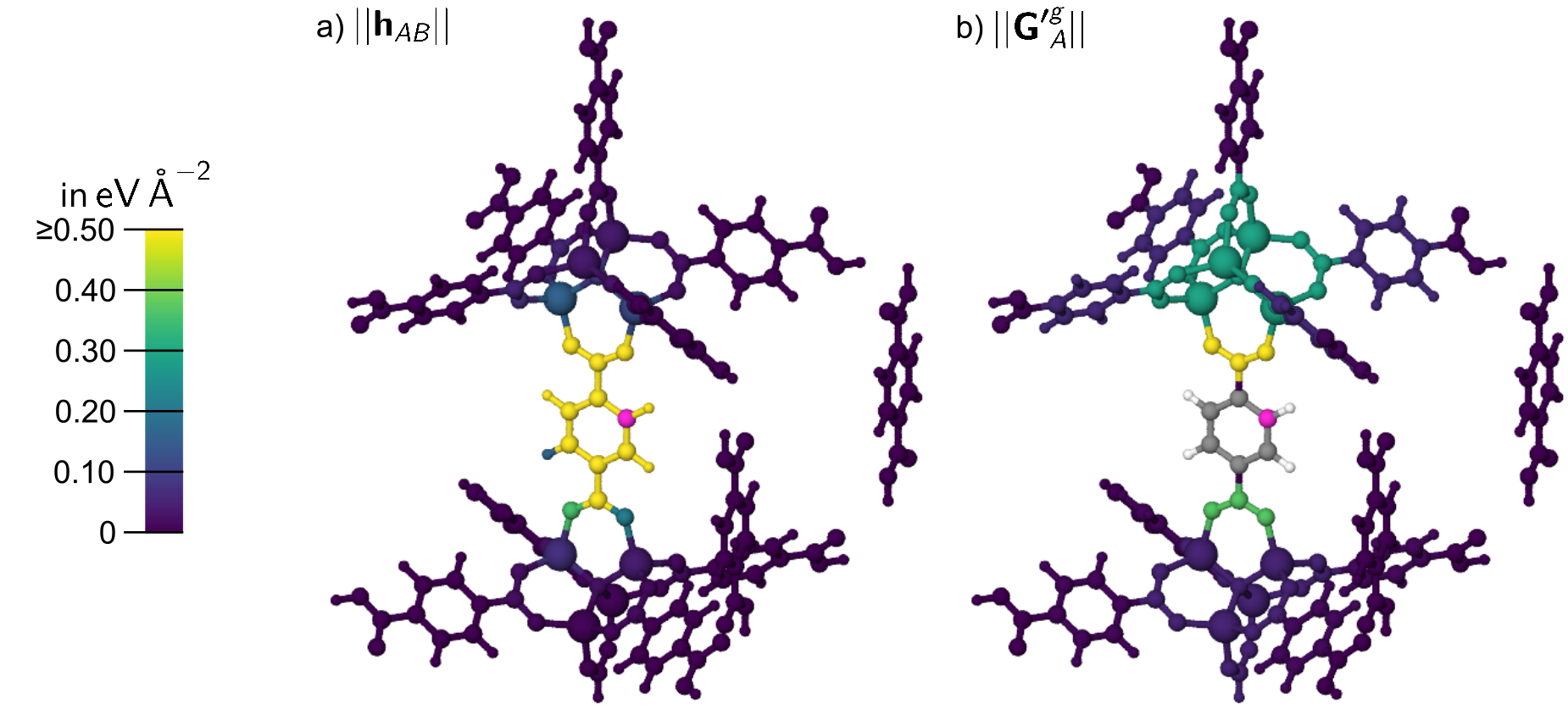}
    \caption{a) Atomic Hessian submatrix norm values $||\mathbf{h}_{AB}||$ and b) effective Hessian group matrix norm values $||\mathbf{G'}_{A}^g||$ in $\text{eV\,\AA{}}^{-2}$ with respect to the central atom $A=\mathrm{C3}$ (magenta) in reference structure $\mathrm{C3_{ref}}$. $||\mathbf{G'}_{A}^g||$ defines the color for the closest atoms of a given group, which in addition also contains all atoms at larger distance. The colors of the smallest possible fragment in b) refer to the chemical elements.}
    \label{fig:FDHessianC3}
\end{figure*}

\begin{figure*}[!hp]
    \centering
    \includegraphics[scale=1.0]{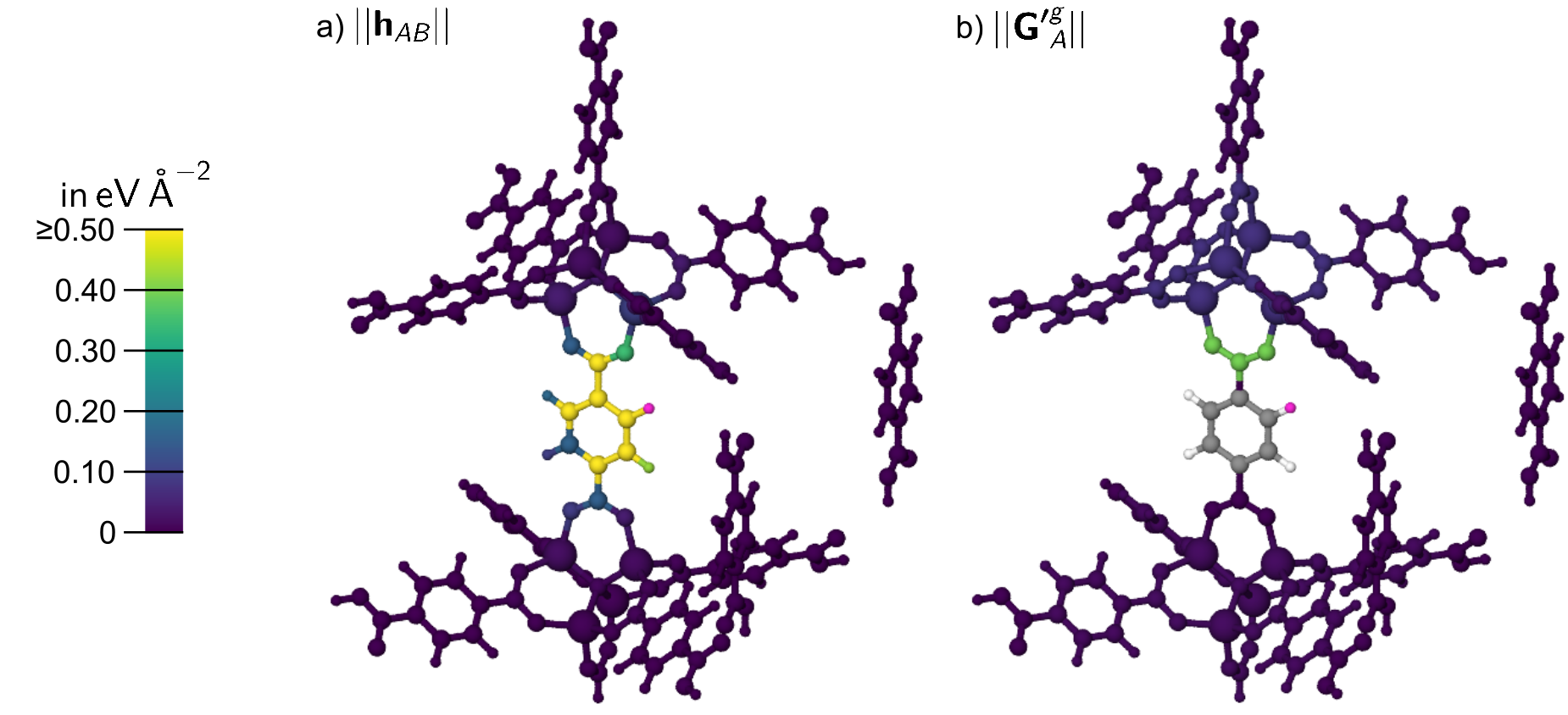}
    \caption{a) Atomic Hessian submatrix norm values $||\mathbf{h}_{AB}||$ and b) effective Hessian group matrix norm values $||\mathbf{G'}_{A}^g||$ in $\text{eV\,\AA{}}^{-2}$ with respect to the central atom $A=\mathrm{H1}$ (magenta) in reference structure $\mathrm{H1_{ref}}$. $||\mathbf{G'}_{A}^g||$ defines the color for the closest atoms of a given group, which in addition also contains all atoms at larger distance. The colors of the smallest possible fragment in b) refer to the chemical elements.}
    \label{fig:FDHessianH1}
\end{figure*}

\begin{table*}[!ht]
     \centering
     \caption{
     Compilation of the force component errors $\Delta f^{Y_g}_{A_{x,y,z}}$ and the total force errors $||\Delta \mathbf{f}^{Y_g}_{A}||$ in $\mathrm{eV\,\text{\AA{}}^{-1}}$ for the reference atoms Zn1, O1, O2, C1, C2, C3, H1 of the reference structures Zn1$_\mathrm{ref}$ (Fig.~\ref{fig:FDHessianZn1}), O1$_\mathrm{ref}$ (Fig.~\ref{fig:FDHessianO1}), O2$_\mathrm{ref}$ (Fig.~\ref{fig:FDHessianO2}), C1$_\mathrm{ref}$ (Fig.~11),
     C2$_\mathrm{ref}$ (Fig.~\ref{fig:FDHessianC2}), C3$_\mathrm{ref}$ (Fig.~\ref{fig:FDHessianC3}) and H1$_\mathrm{ref}$ (Fig.~\ref{fig:FDHessianH1}). Further, the effective Hessian group matrix norm $||\mathbf{G'}_{\mathrm{C1'''}}^{g}||$ is given in $\mathrm{eV\,\text{\AA{}}^{-2}}$. Numbers outside the intended convergence are given in bold.}
     \label{tab:SC_3DFrags}
     \begin{ruledtabular}
     \begin{tabular}{ccccccc}
        $A$/$Y$ & $g$	&	$\Delta f^{Y_g}_{A_\mathrm{x}}$	&	$\Delta f^{Y_g}_{A_\mathrm{y}}$	&	$\Delta f^{Y_g}_{A_\mathrm{z}}$ & $||\Delta\mathbf{f}^{Y_g}_{A}||$	    &	$||\mathbf{G}'^{g}_{A}||$	\\ \hline
Zn1	&	ref	&	0.0000	&	0.0000	&	0.0000	&	0.0000	&	0.00	\\
	&	1	&	0.0016	&	$-0.0018$	&	$-0.0100$	&	0.0103	&	0.27	\\
	&	2	&	0.0065	&	0.0011	&	$-0.0143$	&	0.0158	&	0.26	\\
	&	3	&	$-0.0282$	&	0.0358	&	0.0204	&	0.0499	&	0.23	\\
	&	4	&	0.0005	&	0.0073	&	$-0.0081$	&	0.0109	&	$\mathbf{0.59}$	\\
O1	&	ref	&	0.0000	&	0.0000	&	0.0000	&	0.0000	&	0.00	\\
	&	1	&	0.0001	&	0.0066	&	$-0.0001$	&	0.0066	&	0.27	\\
	&	2	&	0.0001	&	0.0000	&	$-0.0001$	&	0.0001	&	0.22	\\
	&	3	&	0.0000	&	0.0002	&	$-0.0001$	&	0.0002	&	0.33	\\
O2	&	ref	&	0.0000	&	0.0000	&	0.0000	&	0.0000	&	0.00	\\
	&	1	&	$-0.0028$	&	0.0028	&	$-0.0026$	&	0.0048	&	0.02	\\
	&	2	&	0.0220	&	$-0.0233$	&	0.0576	&	0.0659	&	0.09	\\
	&	3	&	0.0268	&	$-0.0281$	&	0.0655	&	0.0762	&	0.25	\\
	&	4	&	0.0143	&	$-0.0155$	&	0.0429	&	0.0478	&	0.22	\\
	&	5	&	0.0058	&	$-0.0070$	&	$-0.0115$	&	0.0146	&	0.23	\\
	&	6	&	$-0.0290$	&	0.0277	&	$-0.0952$	&	0.1033	&	0.15	\\
	&	7	&	$\mathbf{0.1157}$	&	$\mathbf{-0.1168}$	&	$\mathbf{0.5454}$	&	$\mathbf{0.5696}$	&	$\mathbf{3.37}$	\\
C1	&	ref	&	0.0000	&	0.0000	&	0.0000	&	0.0000	&	0.00	\\
	&	1	&	$-0.0018$	&	0.0022	&	0.0165	&	0.0167	&	0.02	\\
	&	2	&	$-0.0019$	&	0.0022	&	$-0.0733$	&	0.0734	&	0.10	\\
	&	3	&	$-0.0019$	&	0.0022	&	$-0.0630$	&	0.0630	&	0.11	\\
	&	4	&	$-0.0013$	&	0.0016	&	$-0.0599$	&	0.0599	&	$\mathbf{0.36}$	\\
	&	5	&	$-0.0024$	&	0.0027	&	0.0172	&	0.0176	&	0.35	\\
	&	6	&	$-0.0018$	&	0.0022	&	$\mathbf{0.1666}$	&	$\mathbf{0.1666}$	&	0.30	\\
	&	7	&	$-0.0016$	&	0.0016	&	$\mathbf{-3.0135}$	&	$\mathbf{3.0135}$	&	$\mathbf{21.83}$	\\
C2	&	ref	&	0.0000	&	0.0000	&	0.0000	&	0.0000	&	0.00	\\
	&	1	&	0.0002	&	$-0.0001$	&	$-0.0115$	&	0.0115	&	0.01	\\
	&	2	&	0.0002	&	$-0.0001$	&	0.0162	&	0.0162	&	0.02	\\
	&	3	&	$-0.0363$	&	0.0364	&	0.0070	&	0.0519	&	0.17	\\
	&	4	&	0.0366	&	$-0.0365$	&	$-0.0068$	&	0.0521	&	$\mathbf{0.36}$	\\
	&	5	&	$\mathbf{0.5746}$	&	$\mathbf{-0.5738}$	&	$-0.0916$	&	$\mathbf{0.8172}$	&	$\mathbf{0.57}$	\\
	&	6	&	$\mathbf{-0.5443}$	&	$\mathbf{0.5442}$	&	$-0.1055$	&	$\mathbf{0.7769}$	&	$\mathbf{0.37}$	\\
	&	7	&	$-0.0002$	&	0.0014	&	$\mathbf{1.8754}$	&	$\mathbf{1.8754}$	&	$\mathbf{20.2}$	\\
C3	&	ref	&	0.0000	&	0.0000	&	0.0000	&	0.0000	&	0.00	\\
	&	1	&	0.0022	&	$-0.0023$	&	0.0017	&	0.0036	&	0.00	\\
	&	2	&	0.0029	&	$-0.0030$	&	0.0015	&	0.0045	&	0.02	\\
	&	3	&	$-0.0252$	&	0.0250	&	$-0.0655$	&	0.0745	&	0.05	\\
	&	4	&	0.0348	&	$-0.0349$	&	0.0668	&	0.0830	&	0.06	\\
	&	5	&	$-0.0236$	&	0.0235	&	$-0.0102$	&	0.0348	&	0.29	\\
	&	6	&	0.0026	&	$-0.0030$	&	0.0006	&	0.0040	&	$\mathbf{0.37}$	\\
	&	7	&	$\mathbf{0.1660}$	&	$\mathbf{-0.1663}$	&	$\mathbf{-0.3542}$	&	$\mathbf{0.4251}$	&	$\mathbf{1.99}$	\\
H1	&	ref	&	0.0000	&	0.0000	&	0.0000	&	0.0000	&	0.00	\\
	&	1	&	$-0.0006$	&	0.0008	&	0.0002	&	0.0010	&	0.01	\\
	&	2	&	0.0003	&	$-0.0001$	&	$-0.0005$	&	0.0006	&	0.02	\\
	&	3	&	0.0010	&	$-0.0007$	&	$-0.0127$	&	0.0128	&	0.02	\\
	&	4	&	$-0.0035$	&	0.0038	&	0.0188	&	0.0195	&	0.03	\\
	&	5	&	$-0.0025$	&	0.0027	&	$-0.0166$	&	0.0170	&	0.01	\\
	&	6	&	$-0.0361$	&	0.0362	&	0.0387	&	0.0641	&	0.07	\\
	&	7	&	$-0.0795$	&	0.0797	&	$-0.0035$	&	0.1126	&	$\mathbf{0.40}$	\\
     \end{tabular}
     \end{ruledtabular}
\end{table*}

\begin{figure*}[!ht]
    \centering
    \includegraphics{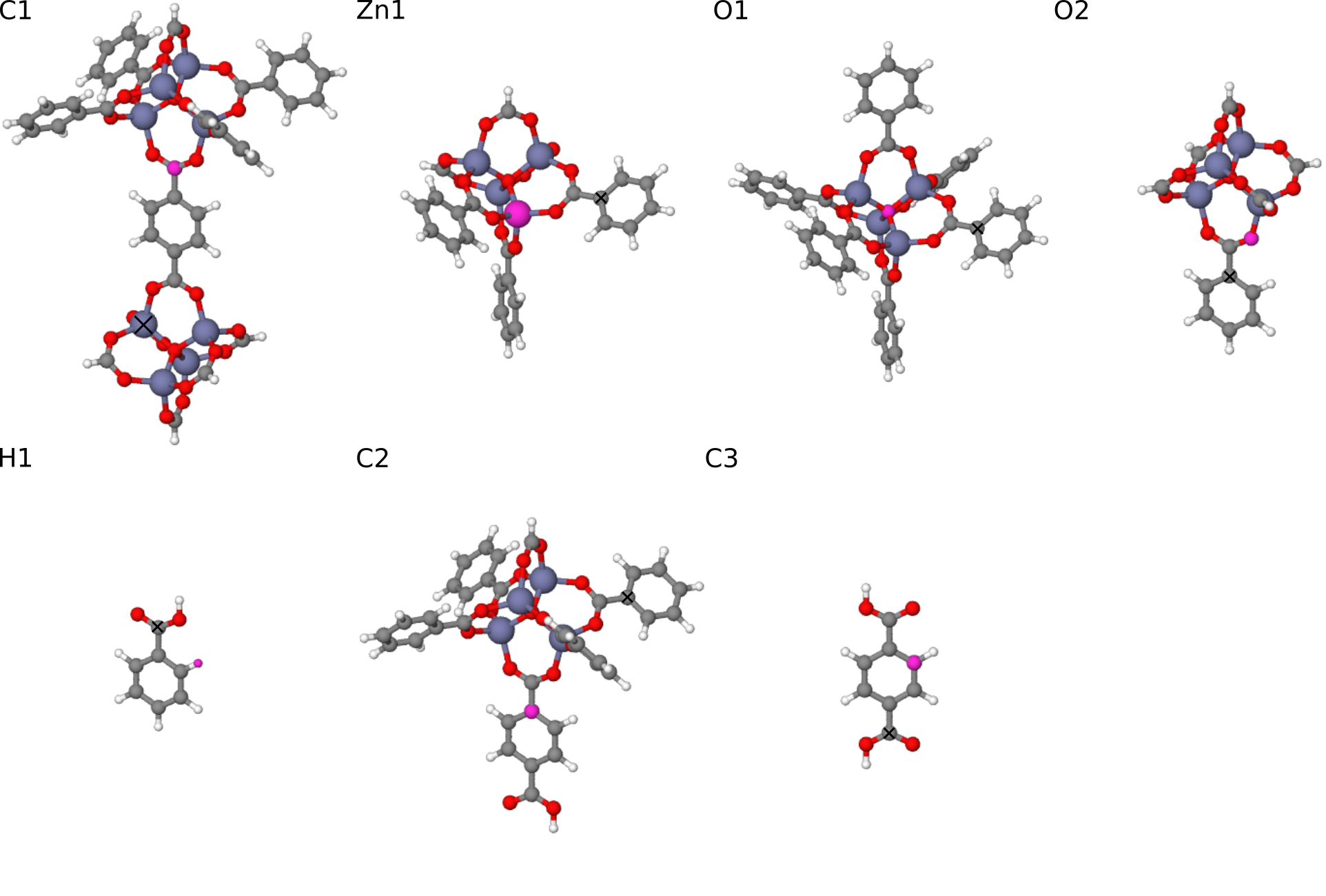}
    \caption{Fragment structures, reduced to the atoms with significant interaction for the chosen convergence criterion of the maximum force component differences $\Delta f_\mathrm{x,y,z}^\mathrm{max} = 0.15\,\mathrm{eV\,\text{\AA{}}^{-1}}$, resulting from FD Hessian analysis for the in-equivalent positions C1, Zn1, O1, O2, H1, C2 and C3 (magenta). The cross marks the atom with the largest $||\mathbf{h}_{AB}||$ of the outermost group $g$.} 
    \label{fig:smallest-frags}
\end{figure*}

\clearpage
\section{Comparison to Bulk MOF-5}\label{sec:DFTbulkMOF5}

\begin{figure}[!ht]
    \centering
    \includegraphics[scale=1.00]{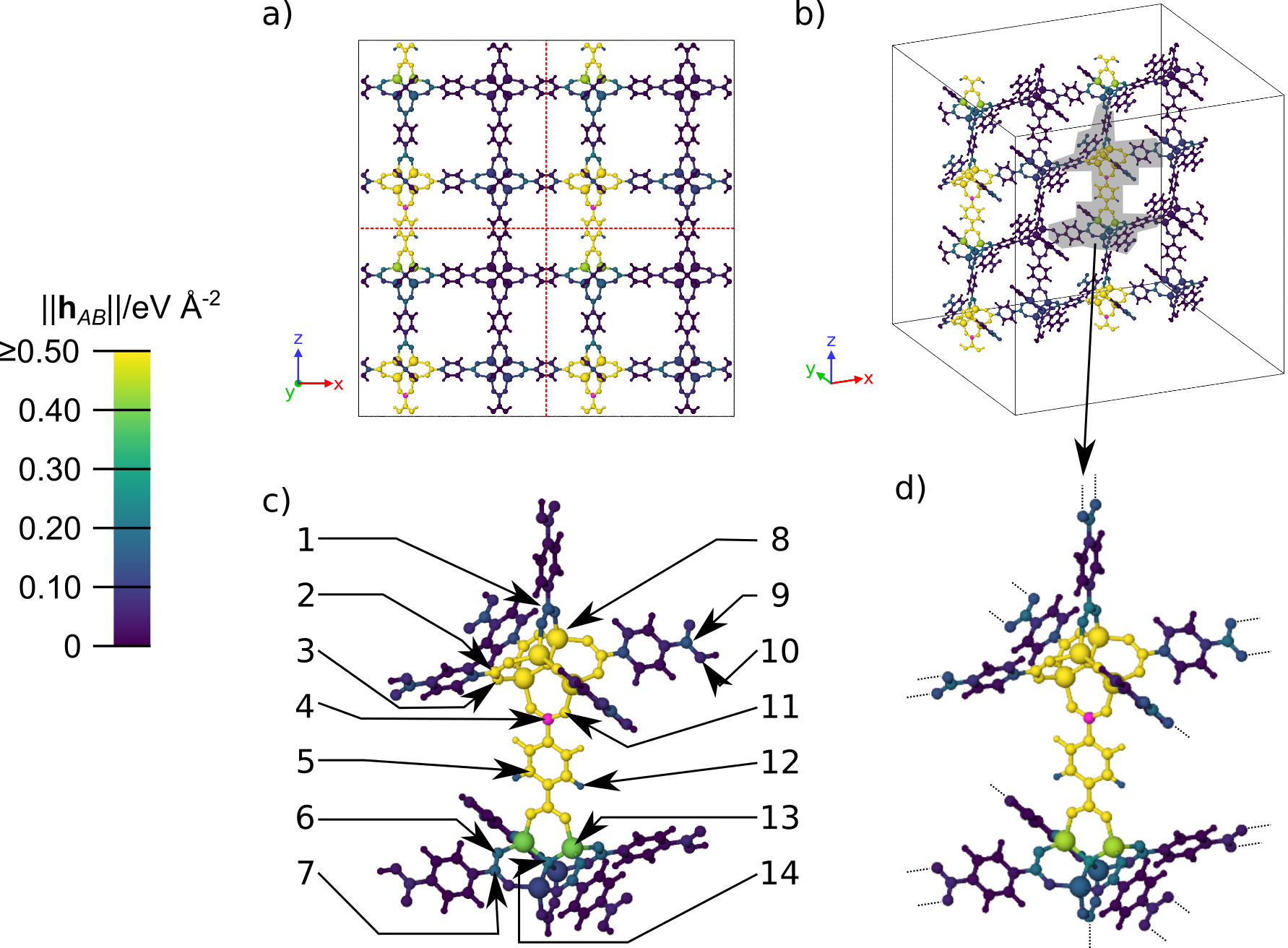}
    \caption{a) In-plane view of the periodic C1 (marked in magenta) atomic Hessian submatrix norm values of the MOF-5 bulk unit cell, shown by a $2 \times 2 \times 2$ replication of the simple unit cell (red dashed lines separate the simple unit cells), b) bent view of a 2D-slab cutout with a grey-marked C1$_\mathrm{ref}$ structure embedded in the periodic MOF-5 environment, c) C1 results of Fig.~11a
    with some marked atoms ($B=1-14$) and d) the grey-marked C1$_\mathrm{ref}$ structure from b) without the surrounding environment and marked broken bonds by the dashed lines.}
    \label{fig:MOF5BulkHessianComparison}
\end{figure}

To validate our approach of using a sufficiently large fragment structure as the Hessian analysis reference structure, we calculated the atomic Hessian submatrix norm values also from the periodic bulk unit cell as shown in Fig.~8a.
In Fig. \ref{fig:MOF5BulkHessianComparison}a the atomic Hessian submatrix norm values, calculated from the $1 \times 1 \times 1$ unit cell, are shown in a $2 \times 2 \times 2$ supercell (Fig. \ref{fig:MOF5BulkHessianComparison}b). All atoms included in C1$_\mathrm{ref}$ are marked by the grey background in Fig. \ref{fig:MOF5BulkHessianComparison}b embedded in the MOF-5 environment. A direct comparison of the C1$_\mathrm{ref}$ fragment (Figs. \ref{fig:MOF5BulkHessianComparison}c and 11a) and periodic results (Fig. \ref{fig:MOF5BulkHessianComparison}d) shows only very small differences (Tab. \ref{tab:CompareC1}).

\begin{table}[!hp]
     \centering
     \begin{ruledtabular}
     \begin{tabular}{ccccc}
        $A$	&	element	&	$||\mathbf{h}_{AB}^\mathrm{C1_{ref}}||$     &	$||\mathbf{h}_{AB}^\mathrm{bulk}||$	& $\Delta||\mathbf{h}_{AB}||$\\ \hline
1	&	C	&	0.143	&	0.218	&	$-0.075$	\\
2	&	C	&	1.170	&	1.205	&	$-0.036$	\\
3	&	O	&	1.218	&	1.265	&	$-0.047$	\\
4	&	C	&	94.885	&	95.788	&	$-0.903$	\\
5	&	C	&	1.098	&	1.106	&	$-0.008$	\\
6	&	O	&	0.185	&	0.202	&	$-0.017$	\\
7	&	C	&	0.172	&	0.193	&	$-0.021$	\\
8	&	Zn	&	0.558	&	0.563	&	$-0.005$	\\
9	&	C	&	0.142	&	0.193	&	$-0.052$	\\
10	&	O	&	0.063	&	0.129	&	$-0.066$	\\
11	&	O	&	42.074	&	42.630	&	$-0.556$	\\
12	&	H	&	0.155	&	0.153	&	0.002	\\
13	&	Zn	&	0.402	&	0.439	&	$-0.037$	\\
14	&	O	&	0.200	&	0.244	&	$-0.045$	\\
     \end{tabular}
     \end{ruledtabular}
     \caption{Atomic Hessian submatrix norm values $||\mathbf{h}_{AB}||$ of the atoms $B=1-14$ marked in Fig. \ref{fig:MOF5BulkHessianComparison}c defined by the reference structures C1$_\mathrm{ref}$ and by the periodic bulk unit cell (Fig.~8a)
     and the differences of these values $\Delta||\mathbf{h}_{AB}||$.}
     \label{tab:CompareC1}
\end{table}

 \bibliography{literature}